\documentclass[12pt]{article}

\usepackage{latexsym,amssymb,euscript}
\usepackage[dvips]{graphicx}
\usepackage{amsmath}
\usepackage{graphbox}
\usepackage{amsfonts}
\usepackage{amssymb, epsfig} 
\usepackage{ulem}

\usepackage{float}
\usepackage{array}

\usepackage[dvipsnames]{xcolor}

\usepackage[utf8]{inputenc}

\usepackage{color,soul}

\begin{document}

\begin{center}
{\Large \textbf{The Fredenhagen-Marcu operator in the gauge-Higgs $Z(2)$ LGT at finite temperature}}

\vspace*{0.6cm}
\textbf{B.~All\'es\footnote{email: alles@pi.infn.it}} \\
\vspace*{0.1cm}
\centerline{\it INFN Sezione di Pisa, Largo Pontecorvo 3, 56127 Pisa, Italy}
\vspace*{0.3cm}
\textbf{O.~Borisenko\footnote{email: oleg@bitp.kyiv.ua}} \\
\vspace*{0.1cm}
\centerline{\it INFN Gruppo Collegato di Cosenza, Arcavacata di Rende, 87036 Cosenza, Italy}
\centerline{\rm and}
\centerline{\it N.N. Bogolyubov Institute for Theoretical Physics,}
\centerline{\it National Academy of Sciences of Ukraine, 03143 Kyiv, Ukraine}
\vspace*{0.3cm}
\textbf{V.~Chelnokov\footnote{email: chelnokov@itp.uni-frankfurt.de}}\\
\vspace*{0.1cm}
\centerline{\it Institut f\"ur Theoretische Physik,}
\centerline{\it Goethe-Universit\"at Frankfurt, 60438 Frankfurt am Main, Germany}
\vspace*{0.3cm}
\textbf{A.~Papa\footnote{email: papa@fis.unical.it}} \\
\vspace*{0.1cm}
\centerline{\it Dipartimento di Fisica, Universit\`a della Calabria}
\centerline{\rm and}
\centerline{\it INFN Gruppo Collegato di Cosenza, Arcavacata di Rende, 87036 Cosenza, Italy}
\end{center}


\begin{abstract}
We explore the possibility to use the Fredenhagen-Marcu operator as a candidate
order parameter of the deconfinement phase transition in gauge-matter 
systems at finite temperature. Concretely, we compute by numerical simulations 
this operator in the $(2+1)$-dimensional $Z(2)$ lattice gauge theory (LGT) with
$Z(2)$ gauge fields coupled to $Z(2)$-valued Higgs fields. While we cannot provide an unambiguous evidence, we present some hints that the Fredenhagen-Marcu operator is capable of distinguishing the deconfinement phase from the Higgs and confinement phases of the theory.
\end{abstract}

\section{Introduction} 

Distinguishing the deconfined phase from the rest of the
phase diagram in gauge theories with dynamical matter fields at finite temperature
remains one of the important unsolved problems in high-energy physics. This problem appears especially relevant in the high-temperature region of QCD, where the deconfined phase is associated with the quark-gluon plasma phase. While the problem
was solved long ago for the pure gauge theory by utilizing
a gauge-invariant order parameter, the Polyakov (or temporal Wilson) loop, to distinguish the deconfined phase from the confined one, the  situation gets much more complicated when dynamical matter fields in the fundamental representation are added to the action. Such fields explicitly break the global center symmetry of the theory and the Polyakov loop ceases to be a good order parameter.  

Two operators able to differentiate between deconfined and other phases at 
{\it zero} temperature were constructed in the past. One is the Preskill-Krauss operator, designed for both Abelian and non-Abelian models with a non-trivial discrete center~\cite{kraus2}. Its possible finite-temperature extension was proposed in~\cite{borisenko_lat97, borisenko_98}. A difficulty with this operator is that, due to its highly non-local nature, it is hard to evaluate it numerically even in Abelian LGTs and there is no analytic tool to calculate the operator in non-Abelian models at small values of the gauge coupling constant, where the appearance of the deconfined phase is expected. 
The second operator is the Fredenhagen-Marcu (FM) operator proposed in~\cite{FM_86}. It measures the ratio of two free energies: the free energy of a staple-like Wilson line squared and the free energy of the Wilson loop of double size in the temporal direction (the precise definition is given below). 
In the limit in which the sizes of the Wilson line and Wilson loop tend to infinity, this
ratio is expected to vanish in the deconfined phase, while it approaches non-zero constant values in the confinement and other phases. 
Thus, the FM operator can test the presence of charged states in the vacuum. 
So far, this operator has been studied mainly in gauge-Higgs LGTs at zero temperature. Rigorous results confirming the expected behavior have been obtained for $4d$ $U(1)$~\cite{fm_4d_u1} and $Z(2)$~\cite{FM_24} gauge-Higgs systems. Numerical studies of the FM operator at zero temperature have been performed in Refs.~\cite{FM_86_MC, z2_gauge_Higgs} for $Z(2)$ models and in~\cite{fm_su2_zero_temp} for the $SU(2)$ model. All results agree with analytic predictions. In a recent paper~\cite{su2_fm_fin_temp} we attempted the computation of the FM operator in the finite temperature $SU(2)$ LGT. The results obtained suggest that, in the above-described large size limit, the FM operator might vanish in the deconfined phase of the theory. However, due to large statistical errors we could not state whether or not this is indeed the case. To get a more reliable answer we have decided to study the (2+1)-dimensional $Z(2)$-invariant gauge-Higgs  model, whose relative simplicity allows to simulate it on larger lattices with bigger statistics. This is the aim of the present paper.
 
The partition function of the model on $\Lambda\in Z^{2+1}$ is defined as
\begin{equation}
\label{gauge_higgs_pf}
Z_{\Lambda}(\beta, \gamma) =  
\sum_{\{ z_n(x)=\pm 1 \} } \ \sum_{\{ s(x)=\pm 1 \} } \ e^{S_G + S_H} \;, 
\end{equation}
where $z_n(x)$ are the gauge fields and $s(x)$ the Higgs fields. The
gauge $S_G$ and the Higgs $S_H$ actions are given by 
\begin{eqnarray}
\label{gauge_action} 
S_G &=& \beta \sum_{x}\sum_{n<m} z_n(x) z_m(x+e_n) z_n(x+e_m) z_m(x)  \ ,   \\ 
\label{higgs_action}   
S_H &=& \gamma \sum_{x}\sum_{n}  \ s(x) z_n(x) s(x+e_n) \ . 
\end{eqnarray} 
Here, $x=(x_1,x_2,t)$, with $t\in [0,N_t-1]$, $x_i\in [0,L-1], i=1,2$, denotes a site on the lattice $\Lambda$. Using the notation $l\equiv (x,n)$ for the link pointing in the direction $n$ from the site $x$, we shall also employ the equivalent notation $z(l)\equiv z_n(x)$. On the other hand, $x+e_n$ is the site beside $x$ towards the direction $n$. 

Periodic boundary conditions are imposed in all directions. The model is invariant under local $Z_{\rm l}(2)$ transformations $\omega(x)$ and global  $Z_{\rm gl}(2)$ transformation $\alpha$:
\begin{eqnarray}
\label{zn_transform}
z_n(x) &\rightarrow& z_n^{\prime}(x) = \omega(x) z_n(x) \omega(x+e_n) \ ,  \\ 
s(x) &\rightarrow& s^{\prime}(x) = \omega(x) s(x) \alpha  \ .
\end{eqnarray}
Thus, the full symmetry group is $Z_{\rm l}(2)\times Z_{\rm gl}(2)$. This model is self-dual in that,
up to a constant, the partition function on the dual lattice maintains the form shown in Eq.(\ref{gauge_higgs_pf}) upon replacing the couplings $\beta,\gamma$ for their dual couplings $\beta_d,\gamma_d$ according to the relations
\begin{eqnarray}
\label{3d_dual_coupl}
\gamma \rightarrow \gamma_d = -\frac{1}{2} \ln \tanh \beta \ ,  \qquad
\beta \rightarrow \beta_d = -\frac{1}{2} \ln \tanh \gamma \ .
\end{eqnarray} 

This paper is organized as follows. In Sec.~2 we define the FM operator and discuss some simple analytic arguments to justify why we expect that its large size limit vanishes in the deconfined phase. Sec.~3 describes our numerical results and in Sec.~4 we present our conclusions.  

\section{The FM operator}
\label{fm_def_analytic}

The authors in~\cite{FM_86} define the FM operator as follows.
Let ${\cal{C}}$ be a closed rectangular loop of size $T\times R$, with $T$ and $R$ the lengths along respectively the temporal and one spatial direction. We denote by $W({\cal{C}})$ the corresponding Wilson loop,
\begin{equation}
\label{wilson_loop_def}
W({\cal{C}}) = \prod_{l\in \cal{C}} \ z(l) \ . 
\end{equation}
By cutting $W({\cal C})$ into two halves along the temporal sides, we obtain a $\sqcap$- and a $\sqcup$-shaped objects. Let us take one of them, for instance
$\sqcap$, and call it ${\cal L}_{xy}$, where $x,y$ are the end points. Hence, the horizontal side of $\sqcap$ is $R$ lattice spacings long in the spatial direction, while the vertical sides run along the temporal direction and have lengths $T/2$. We then consider a gauge-invariant Wilson line, denoted by $V({\cal L}_{xy})$, and obtained by multiplying ${\cal L}_{xy}$ with the Higgs fields at the end points,
\begin{equation}
\label{wilson_line_def}
V({\cal{L}}_{xy}) =  s(x) \ \Big( \prod_{l\in {\cal{L}}_{xy}} \ z(l) \Big) \ s(y) \ . 
\end{equation}
The FM operator $\rho$ is defined as the value of the ratio
\begin{equation}
\label{FMoperator_def}
H(R,T) = 
\frac{\langle V({\cal{L}}_{xy}) \rangle^2}{\langle W({\cal{C}}) \rangle} \ ,
\end{equation}
in the large distance limit, {\it viz.}
\begin{equation}
\label{FMoperator_defBIS}
\rho \equiv \lim_{R\to\infty} \ H(R,T) \ .
\end{equation} 
However, we will often simplify the language by employing ``FM operator" to refer also to the ratio~(\ref{FMoperator_def}) itself.

Note that the above definition differs from the conventional one where one usually takes $T=R$ before the limit in~(\ref{FMoperator_defBIS}), so that in the limit $R\to\infty$ both temporal and spatial sizes of the Wilson loop and line diverge. 
At finite temperature the size $T$ cannot exceed the temporal lattice extent $N_t$, which is less than the spatial size $L$. 
The operator obtained by setting $T$ to a constant value will be referred to as
temporal FM operator. In the present work we compute both the temporal and the spatial FM operators. 
The latter is defined by taking spatial Wilson loops and spatial Wilson lines in Eq.(\ref{FMoperator_def}) and imposing the standard constraint $T=R$.

It is not obvious that the large distance limit of the temporal FM operator vanishes in the deconfined phase if only $R$ goes to infinity while $T$ remains fixed. In Ref.~\cite{su2_fm_fin_temp} we discussed simple non-rigorous arguments which prove why this can be the case in the $SU(2)$ gauge-Higgs theory at finite temperature. Below we briefly outline those arguments for the $Z(2)$ theory. 
It is also possible to prove rigorously that the large $R$ limit of any FM operator vanishes in the deconfined phase of the $Z(2)$ theory, if $T$ is fixed and the limit is taken as in (\ref{FMoperator_defBIS})~\cite{priv_com}. 

Consider the region of small Higgs coupling $\gamma\ll 1$. Here the system can be in two phases, confined or deconfined, depending on the value of the gauge coupling. In the confined phase, corresponding to small $\beta$,  
the Wilson loop in the pure gauge theory decays with the area law 
$\langle W({\cal{C}}) \rangle_G\approx (\tanh\beta)^{RT}$ and the leading 
perimeter-law contribution comes from the Higgs action. 
Instead, for the Wilson line one gets
\begin{eqnarray}
\label{wilson line_z2_conf}
\langle V({\cal{L}}_{xy}) \rangle  &=& (\tanh\gamma)^{R + T} 
\left [ 1 + (2R+2T-2) \tanh^2\gamma \tanh\beta + \cdots  \right ]  \\
&& +\sum_{{\cal{M}}_{xy}} \ 
( \tanh\gamma )^{| {\cal{M}}_{xy} |} \ 
\langle W({\cal{L}}_{xy} \cup {\cal{M}}_{xy}) \rangle_G   \ ,  \nonumber 
\end{eqnarray}
where the sum runs over all paths ${\cal M}_{xy}$ connecting points $x$ and $y$, while the Wilson loop $W({\cal{L}}_{xy} \cup {\cal{M}}_{xy})$ is constructed by closing the end points of ${\cal{L}}_{xy}$ with the path ${\cal{M}}_{xy}$. The suffix $G$ in $\langle \cdots \rangle_G$ indicates that the expectation value is extracted in the pure gauge theory. Then, for any FM operator we find 
\begin{equation}
\label{FM_z2_conf} 
H(R,T) \approx \frac{\left ( (\tanh\gamma)^{R + T} + \cdots + (\tanh\gamma)^{R} \ (\tanh\beta)^{\frac{RT}{2}} + \cdots \right )^2}{(\tanh\gamma)^{2R + 2T} 
+ \cdots + (\tanh\beta)^{RT} + \cdots }  \ . 
\end{equation}
The ratio goes to a constant in the limit $R\to\infty$, if $T$ is fixed to a value larger than~2.

In the region of large $\beta$ both temporal and spatial Wilson loops in the pure gauge theory decay with the perimeter law 
$\langle W({\cal{C}}) \rangle_G\approx \exp\left ( -m(\beta) P({\cal{C}}) \right ) $.  One gets
\begin{equation}
\label{wilson loop_z2_deconf}
\langle W({\cal{C}}) \rangle = \exp\left ( -m(\beta) (2R + 2T) \right ) 
+ (\tanh\gamma)^{2R + 2T} \ ,   
\end{equation}
\begin{equation} 
\label{wilson line_z2_deconf}
\langle V({\cal{L}}_{xy}) \rangle  = (\tanh\gamma)^{R + T} + 
\sum_{{\cal{M}}_{xy}} \ 
\left ( \tanh\gamma \right )^{| {\cal{M}}_{xy} |} \ 
\exp\left ( -m(\beta) (R + T + | {\cal{M}}_{xy} | ) \right ) \ , 
\end{equation}
where the constant $m(\beta)\sim{\cal{O}}(e^{-\beta})$. The leading contribution comes from the term $| {\cal{M}}_{xy} |=R$ and gives 
\begin{equation}
\label{FM_z2_deconf_1} 
H(R,T) \approx (\tanh\gamma)^{2R} \ \frac{\left ( (\tanh\gamma)^T +  e^{-m(\beta) (2R + T)}  \right )^2}{(\tanh\gamma)^{2R + 2T} + e^{-m(\beta) (2R + 2T)}} \ . 
\end{equation}
There is a competition between the two terms in the denominator: the first term arises due to screening from the matter fields, while the second term arises from the gluonic screening. If $\beta$ is sufficiently large, the gluonic screening wins 
and we obtain  
\begin{equation}
\label{FM_z2_deconf_final} 
H(R,T) \approx  \left ( \tanh\gamma \ e^{m(\beta)} \right )^{2R + 2T} \ .
\end{equation}
Even if $T$ is fixed, $H(R,T)\to 0$ in the limit $R\to\infty$.

\section{Numerical results}
\label{num_results}

{\it The phase diagram and universality class}.
The phase structure of the $Z(2)$ gauge-Higgs LGT is well known  both at zero and finite temperatures~\cite{pisa_rev_25}. It includes three phases: confinement, deconfinement, and Higgs. The confinement phase at small gauge and Higgs couplings is separated from the Higgs phase at large $\gamma$ by a smooth transition where the free energy and its derivatives remain analytic functions, in agreement with~\cite{seiler_78, fradkin_shenker}. This transition can, however, be described in analogy to spin-glass models by using an appropriate generalization of the overlap operator, see Ref~\cite{greensite_22_overlap} for a review of this approach. Such overlap operator was computed numerically for the $Z(2)$ model at zero temperature in~\cite{z2_gauge_Higgs}. 

The deconfinement phase is separated from the confinement and Higgs phases by two lines representing thermodynamic transitions, which are in the $3d$ Ising universality class. These two lines merge in a multicritical point and continue as one line of first order phase transitions for a short interval. The end point of this first order transition is again in the Ising universality class.

\begin{figure}[H]
\centering
\includegraphics[width=0.32\linewidth,clip]
{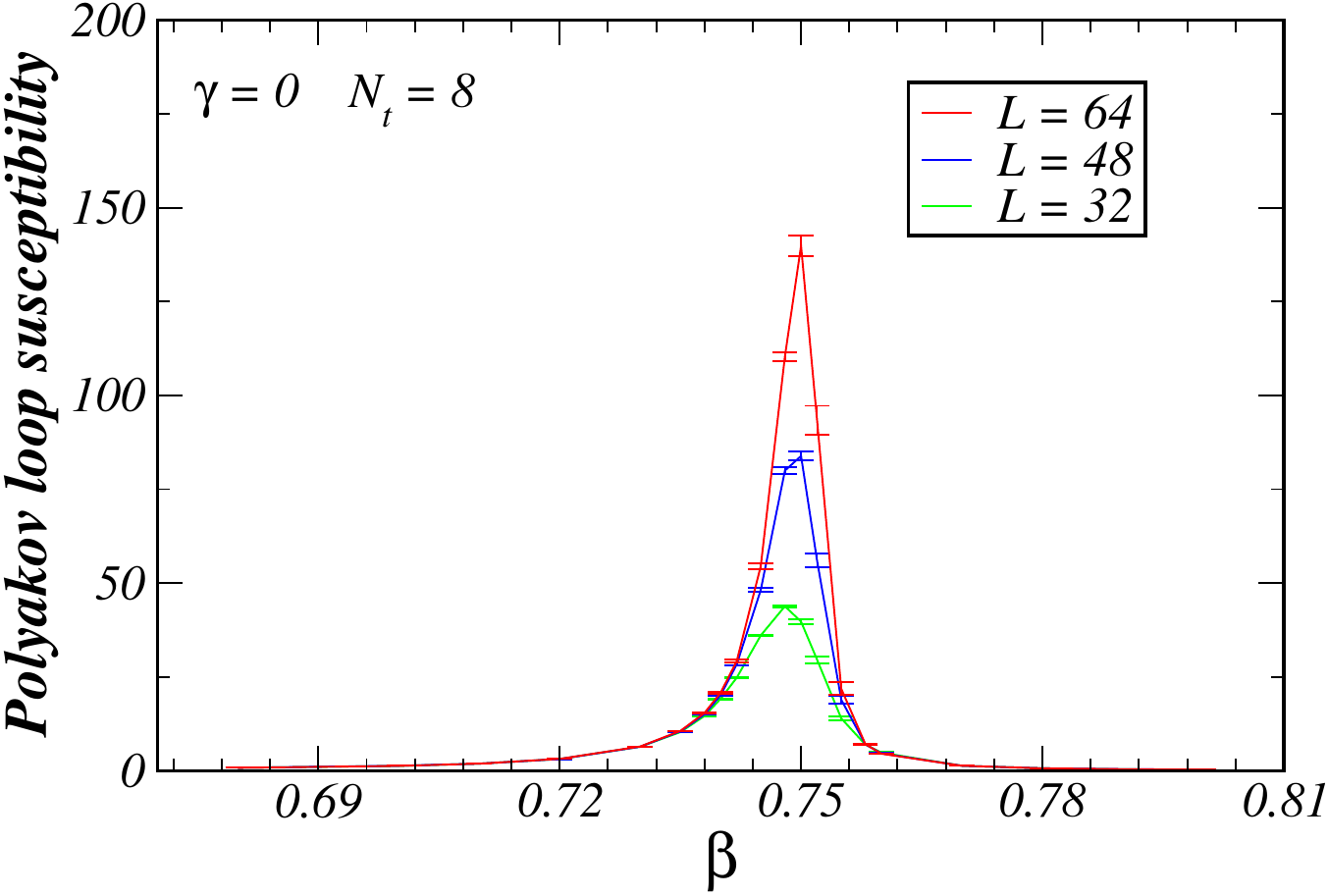}
\includegraphics[width=0.32\linewidth,clip]
{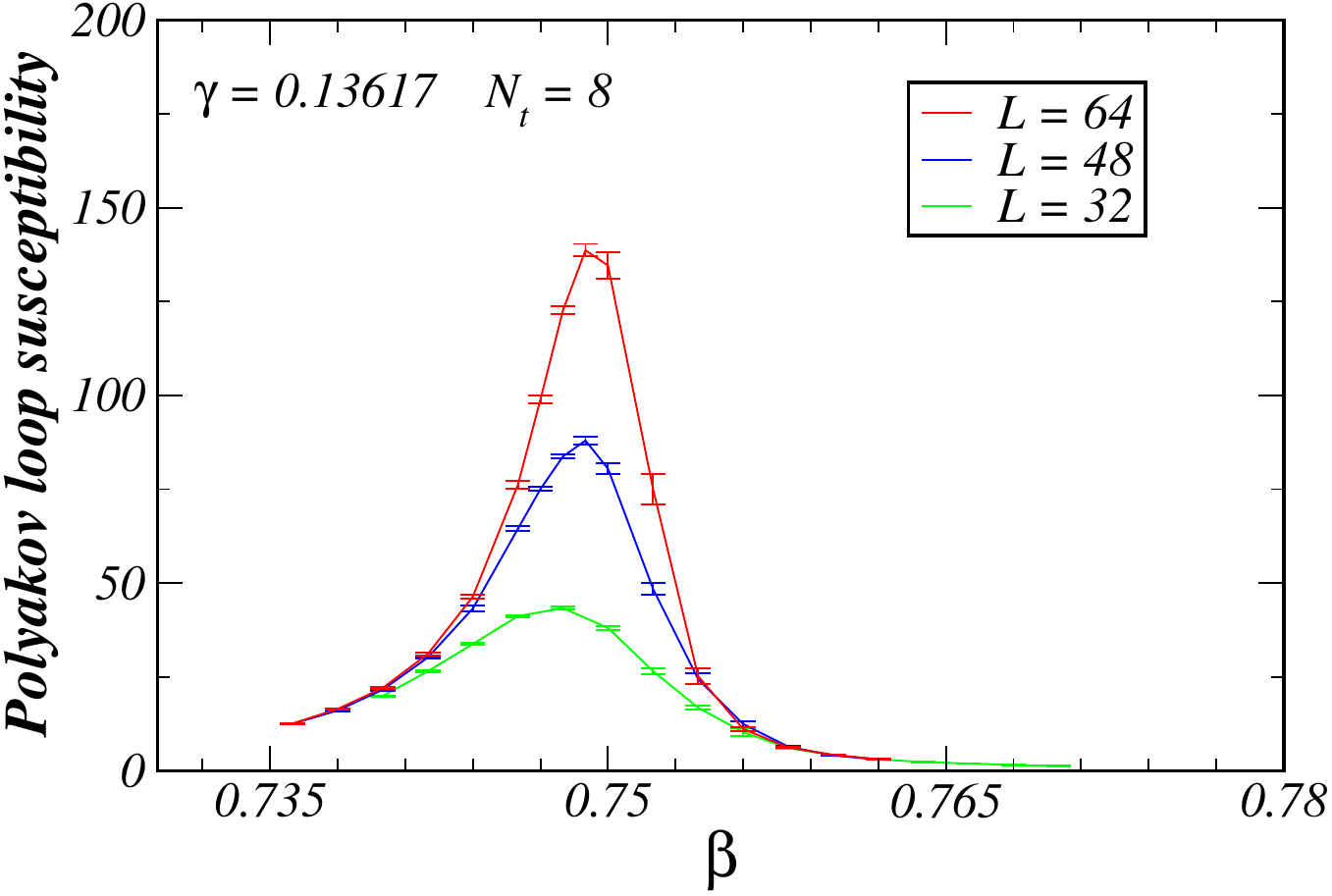}
\includegraphics[width=0.32\linewidth,clip]
{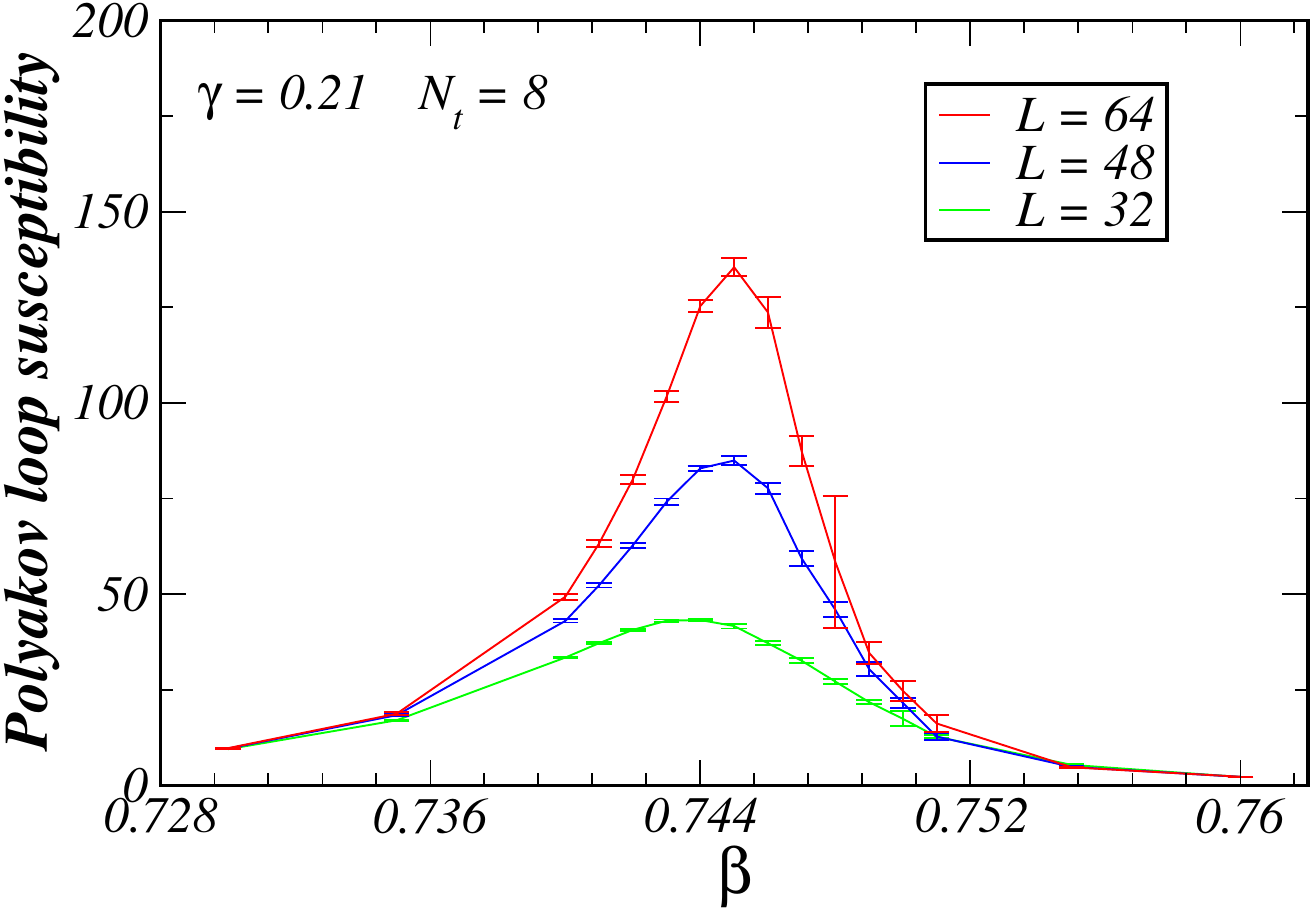}

\includegraphics[width=0.32\linewidth,clip]
{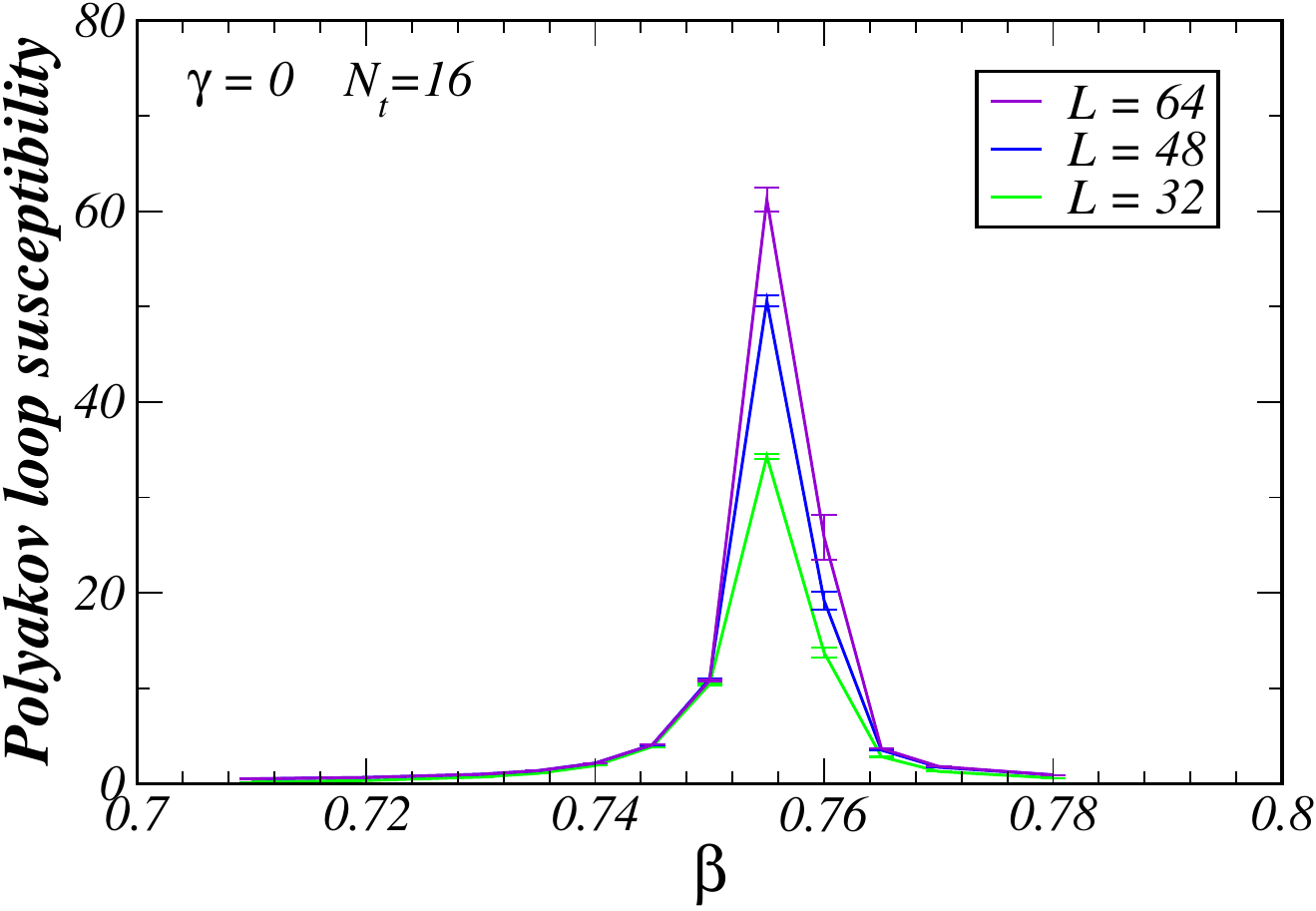}
\includegraphics[width=0.32\linewidth,clip]
{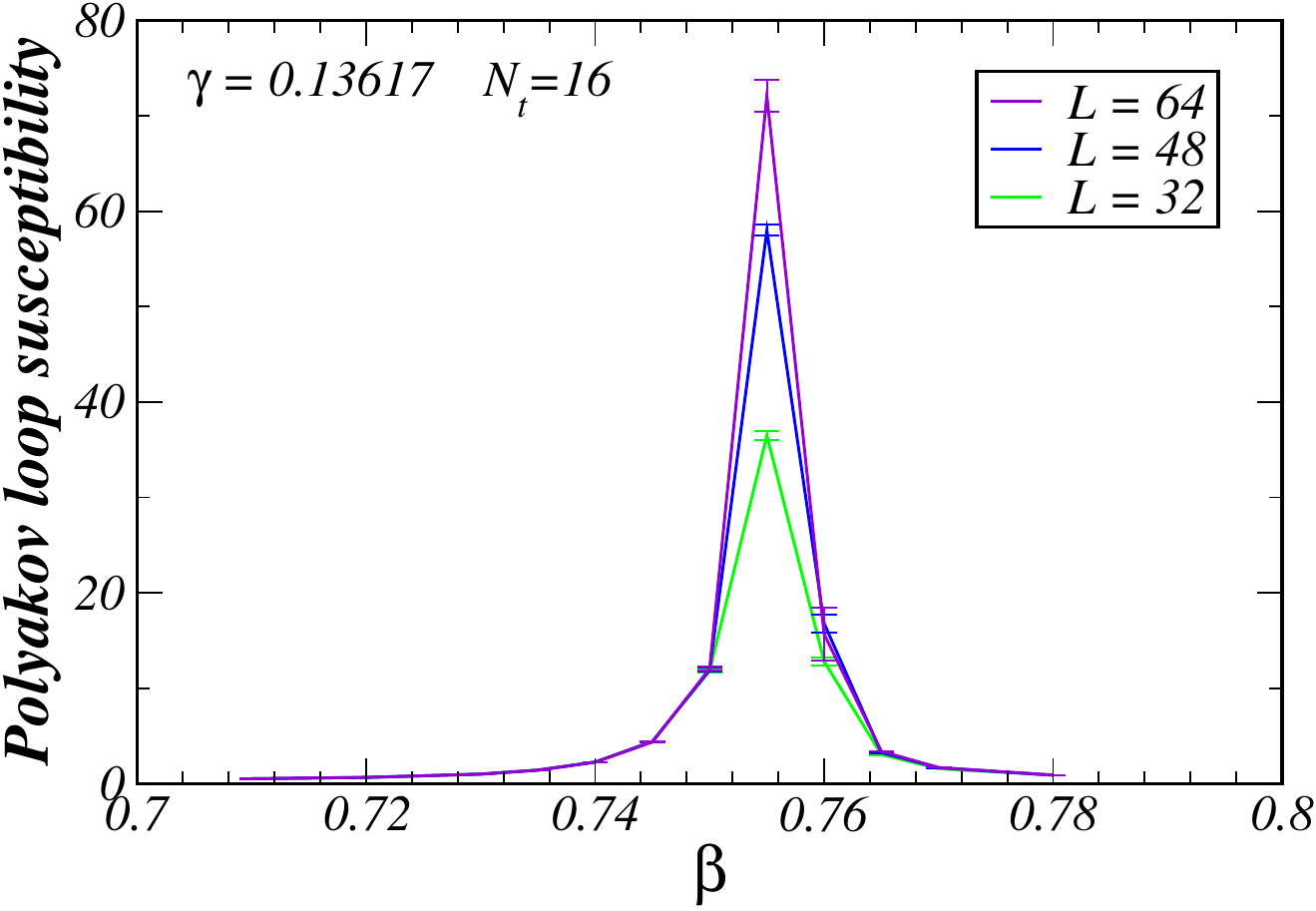}
\includegraphics[width=0.32\linewidth,clip]
{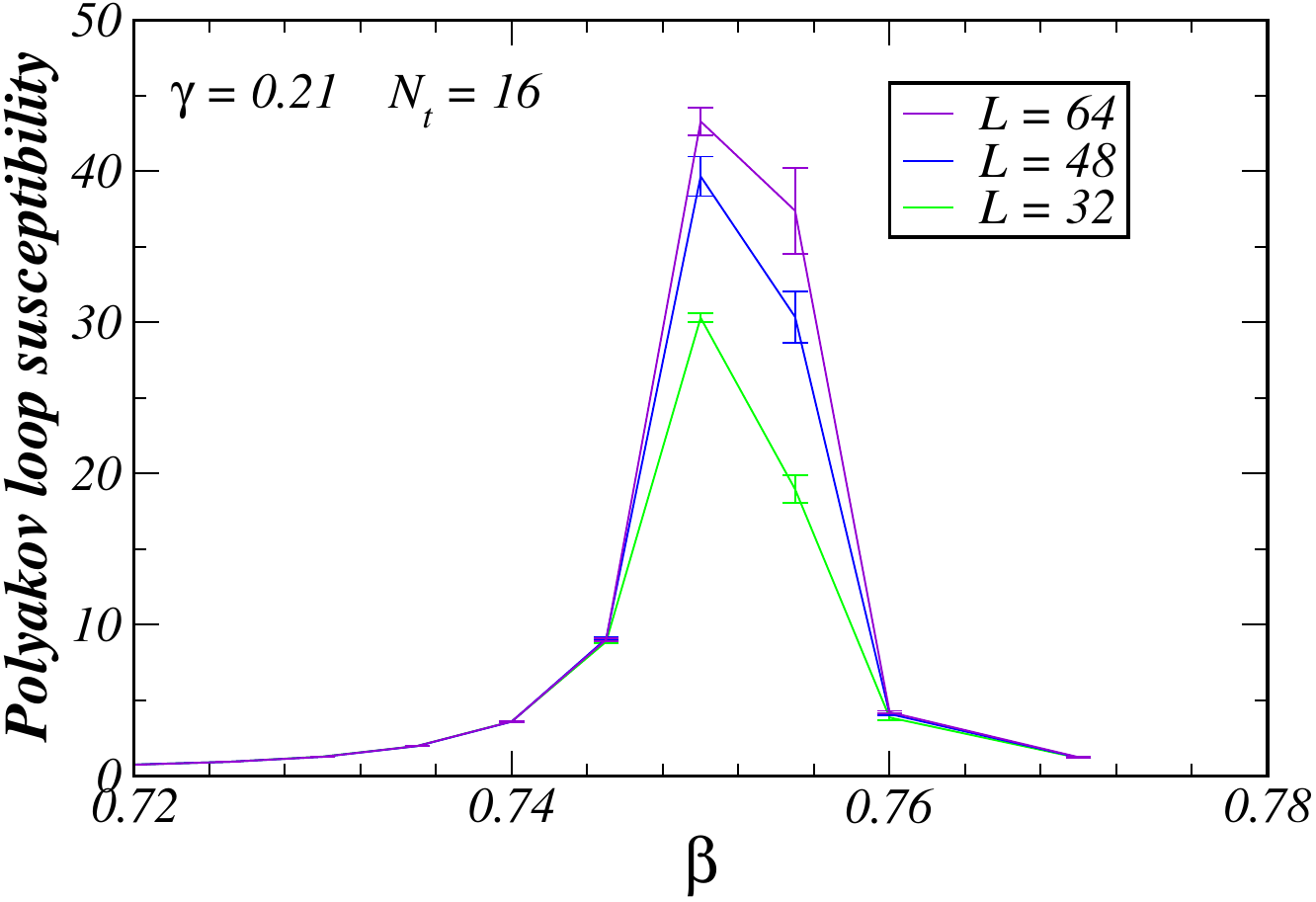}
\caption{The Polyakov loop susceptibility as a function of $\beta$ at fixed values of $\gamma$ in the region of the second order phase transition for several spatial and temporal sizes.}
\label{fig:pl_susc_term_tr}
\end{figure}

\begin{figure}[H]
\centering
\includegraphics[width=0.32\linewidth,clip]
{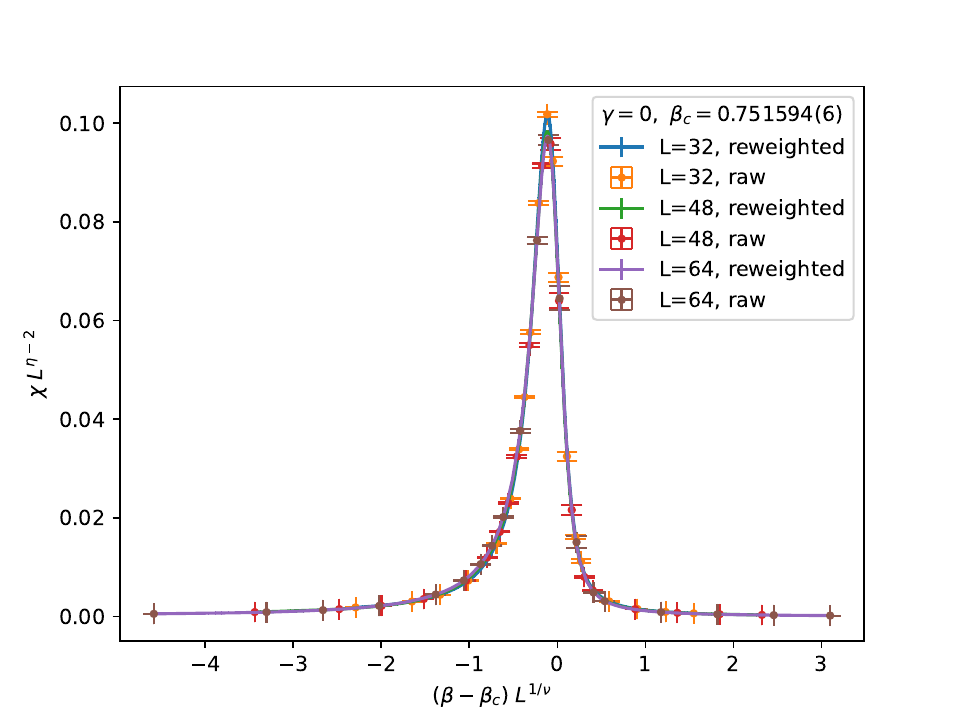}
\includegraphics[width=0.32\linewidth,clip]
{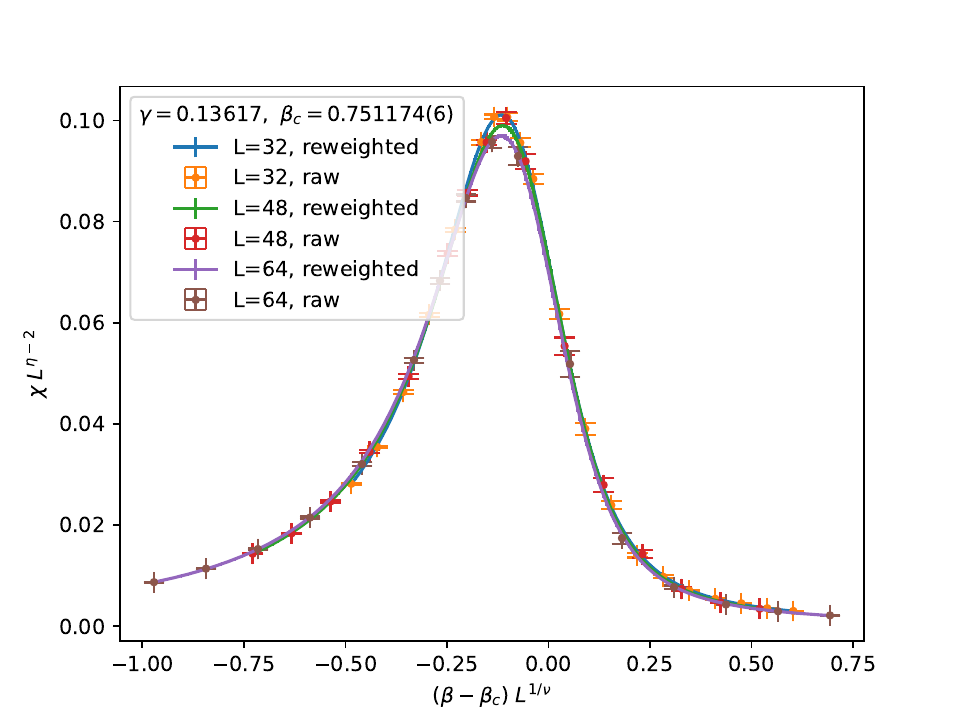}
\includegraphics[width=0.32\linewidth,clip]
{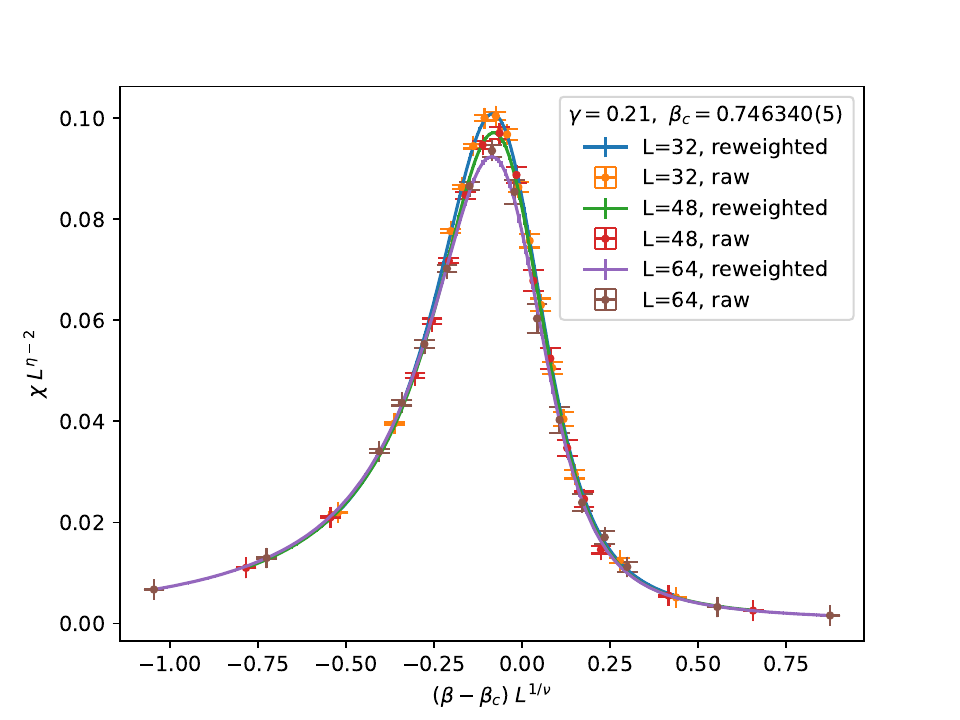}

\includegraphics[width=0.32\linewidth,clip]
{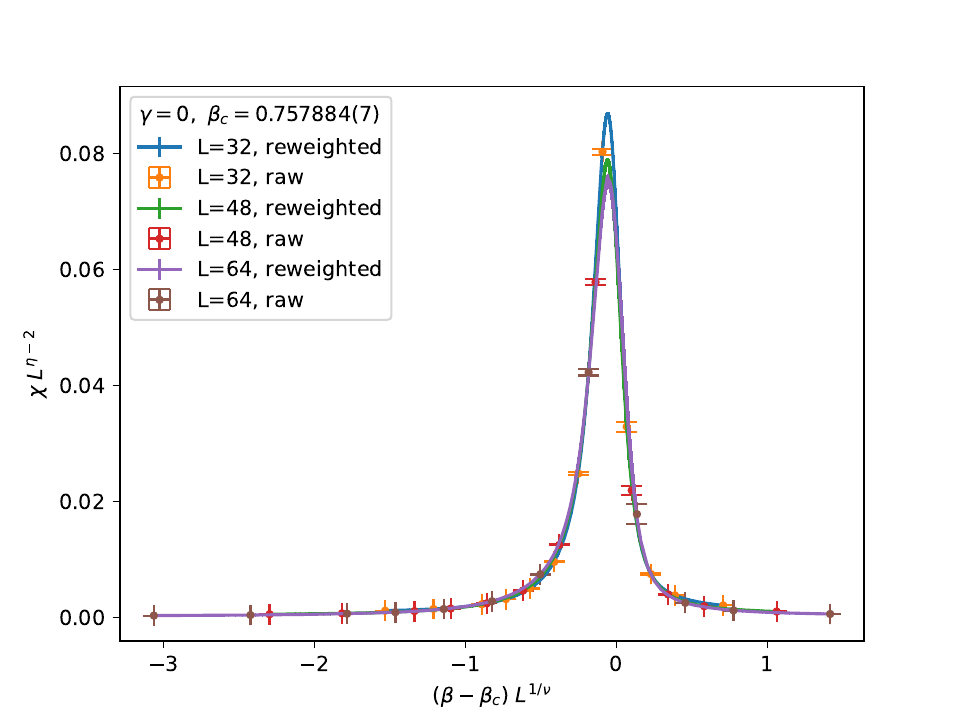}
\includegraphics[width=0.32\linewidth,clip]
{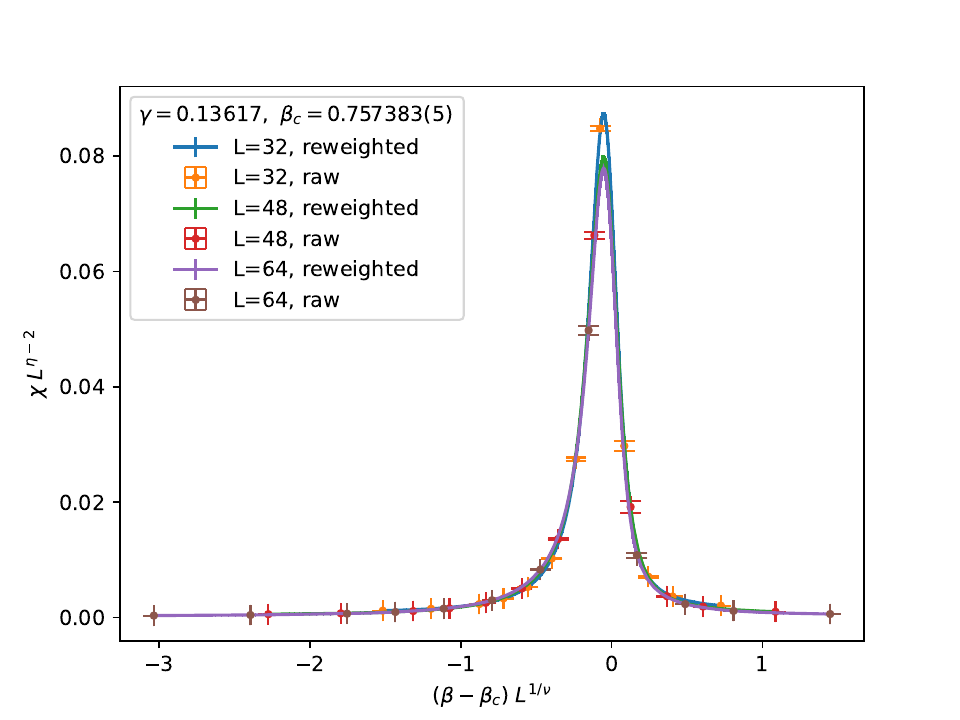}
\includegraphics[width=0.32\linewidth,clip]
{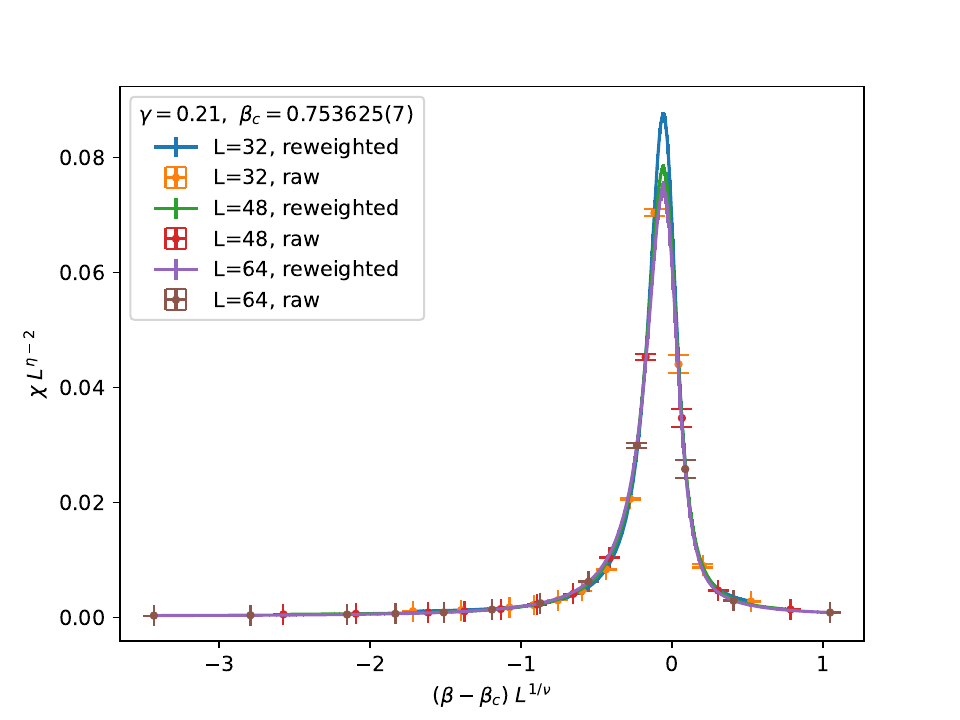}
\caption{Collapse of the Polyakov loop susceptibility. Data obtained on $N_t=8$ (top) and $N_t = 16$ (bottom) lattices for several spatial sizes $L$. Critical indices are fixed to the values in $2d$ Ising model: $\eta=0.25$ and $\nu=1$. ``Raw'' data refers to the points obtained directly from Monte-Carlo simulations, while ``reweighted'' data are obtained using the multihistogram reweighting of raw data to intermediate values of gauge coupling $\beta$.}
\label{fig:chi_collapse}
\end{figure}

\begin{figure}[H]
\centering
\includegraphics[width=0.32\linewidth,clip]
{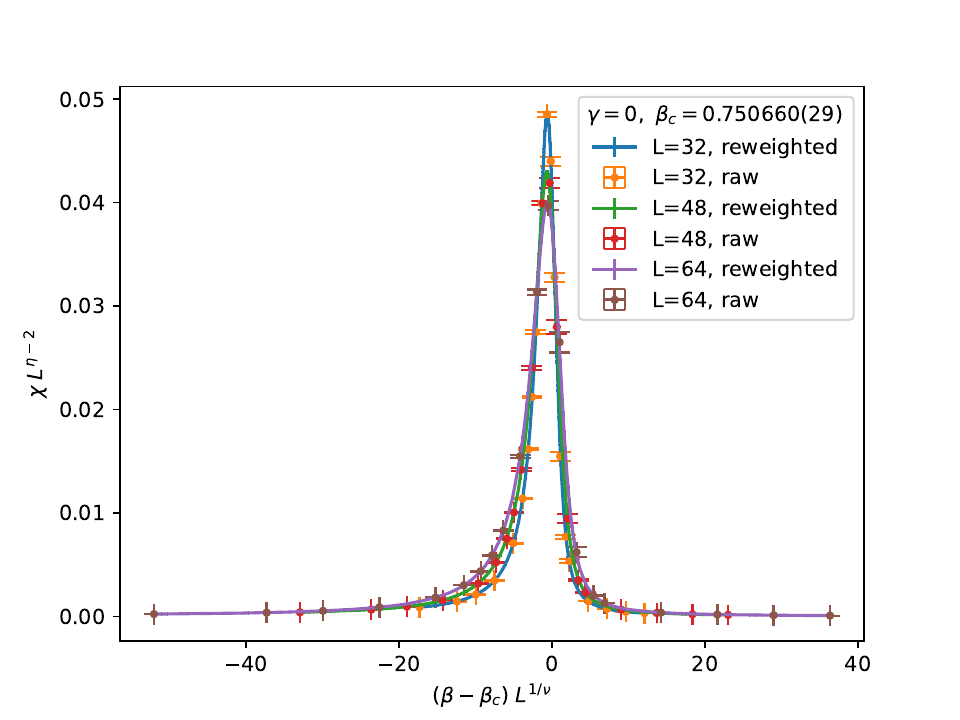}
\includegraphics[width=0.32\linewidth,clip]
{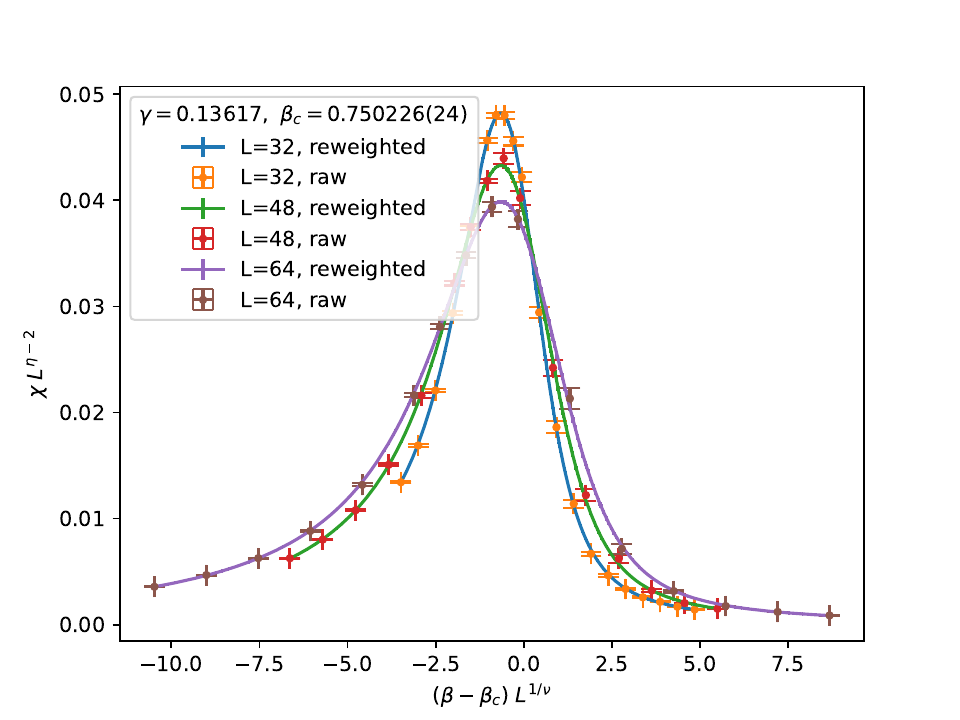}
\includegraphics[width=0.32\linewidth,clip]
{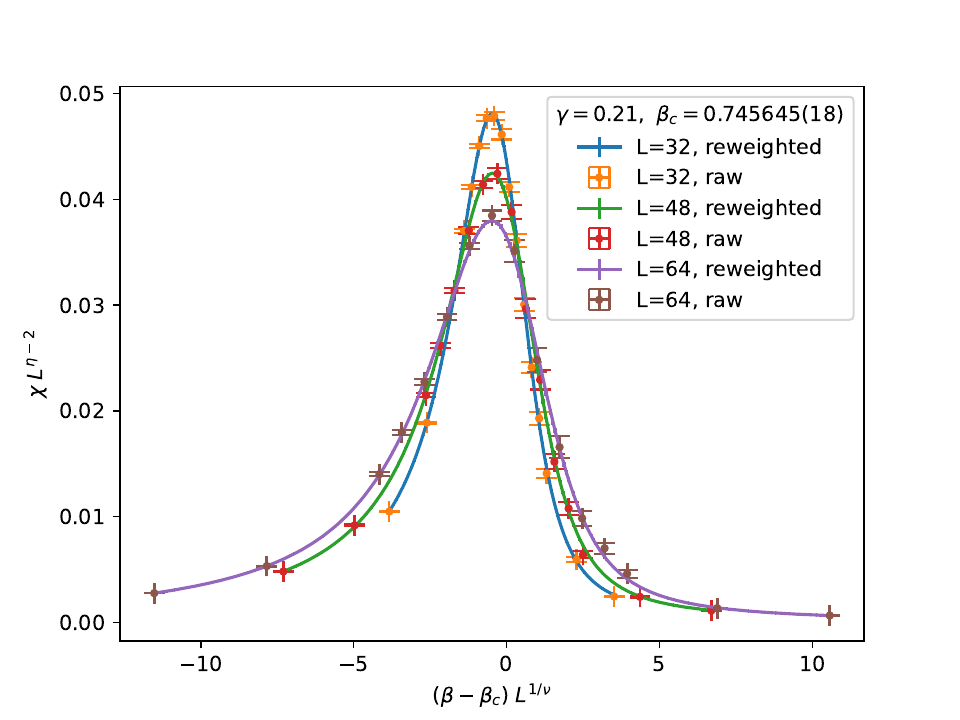}

\includegraphics[width=0.32\linewidth,clip]
{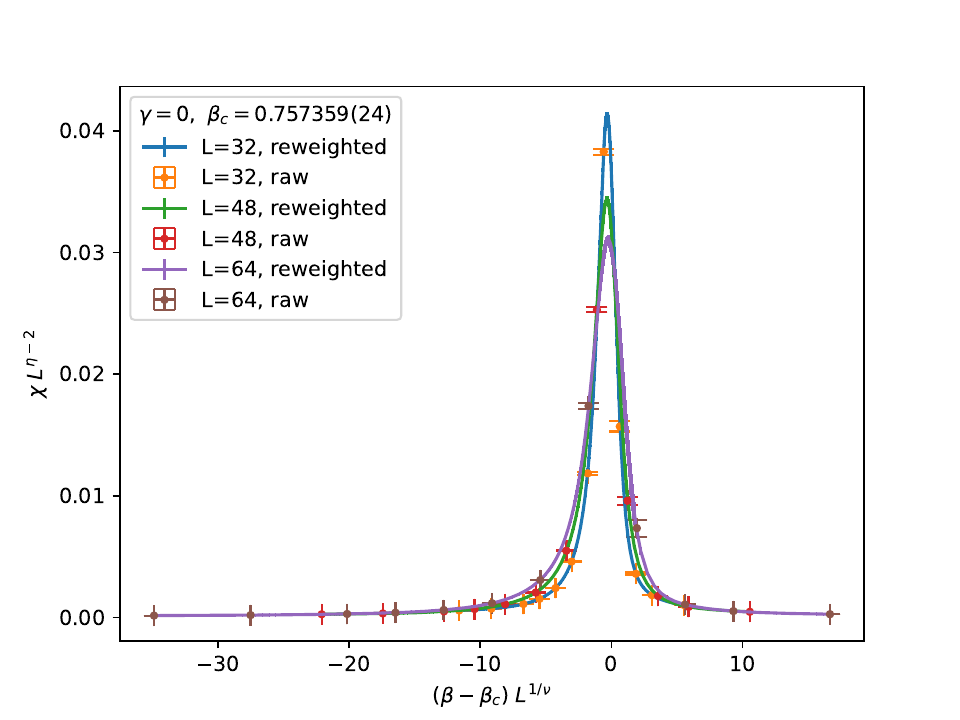}
\includegraphics[width=0.32\linewidth,clip]
{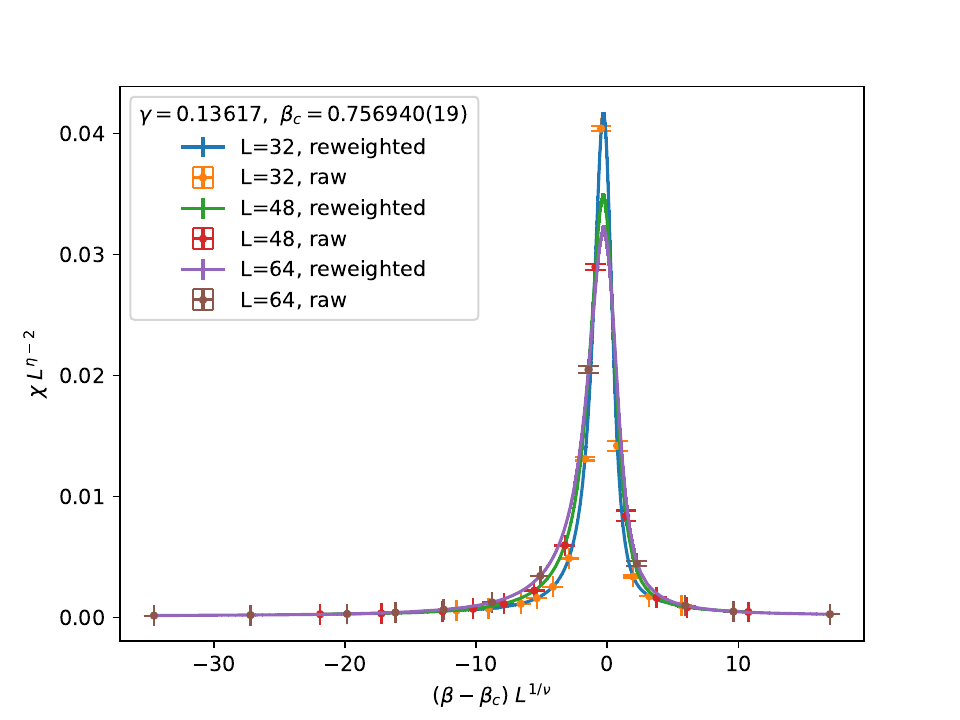}
\includegraphics[width=0.32\linewidth,clip]
{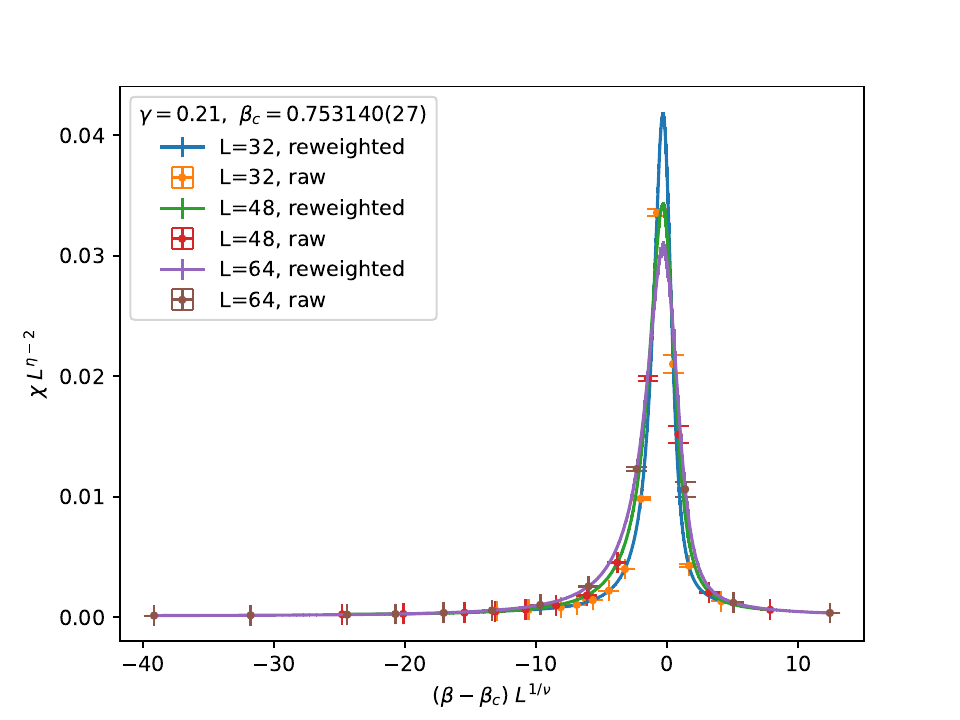}
\caption{Collapse of the Polyakov loop susceptibility. Data obtained on $N_t=8$ (top) and $N_t = 16$ (bottom) lattices for several spatial sizes $L$. Critical indices are fixed to the values in $3d$ Ising model: $\eta\approx0.03629$ and $\nu\approx0.63$. The meanings of the labels are the same as in Fig.\ref{fig:chi_collapse}.}
\label{fig:chi_collapse_3dIsing}
\end{figure}

The phase structure just described is expected to stay essentially the same at finite temperature. The general phase diagram can be established by computing the Polyakov loop susceptibility. In Fig.\ref{fig:pl_susc_term_tr} we show this quantity in the vicinity of the confinement-deconfinement phase transition. Here, we fix the Higgs coupling at three different values
($\gamma=0$, 0.13167 and 0.21), and compute the Polyakov loop at various values of $\beta$ on a lattice with temporal extents $N_t$=8 and $N_t=16$ and several lattice spatial sizes $L=32, 48, 64$. 
The peak of the susceptibility is clearly seen and grows with the lattice spatial size indicating a thermodynamic transition. 
To determine the precise location of the peak we employ multihistogram reweighting~\cite{multihist} to obtain the dependence of the susceptibility $\chi$ on 
$\beta$ in the vicinity of the simulated points. 
For each spatial size $L$, the peak position $\beta_\mathrm{pc}(L)$ is identified from the condition $\frac{\partial \chi}{\partial \beta} = 0$.
These pseudocritical couplings are then extrapolated to the thermodynamic limit using the finite-size scaling relation 
\begin{equation}
\label{FSS}
\beta_\mathrm{pc}(L) = \beta_\mathrm{c} - A L^{-\frac{1}{\nu}} \ .
\end{equation}
Once $\beta_\mathrm{c}$ is known, we can test universality by plotting the rescaled Polyakov loop susceptibility $\chi L^{\eta-2}$ as a function of the rescaled coupling $(\beta-\beta_\mathrm{c}) L^{\frac{1}{\nu}}$. 
Fig.\ref{fig:chi_collapse} shows this scaling collapse using the $2d$ Ising critical exponents $\nu = 1$, $\eta = 1/4$. The collapse is good except for a narrow region near the peak for the smallest lattice size ($L=32$), suggesting 
that the transition belongs to the $2d$ Ising universality class. 
For comparison, we repeated the same analysis using the $3d$ Ising critical exponents $\nu \approx 0.6299709$, $\eta \approx 0.036297612$, with the result  shown in Fig.\ref{fig:chi_collapse_3dIsing}. 
In this case the collapse fails both for the peak height, and width. 
In this way we locate the (vertical) critical line between confinement and deconfinement phases. The (horizontal) critical line between deconfinement and Higgs phases can be calculated from the duality relations (\ref{3d_dual_coupl}). 

\begin{figure}[H]
\centering
\includegraphics[width=0.48\linewidth,clip]
{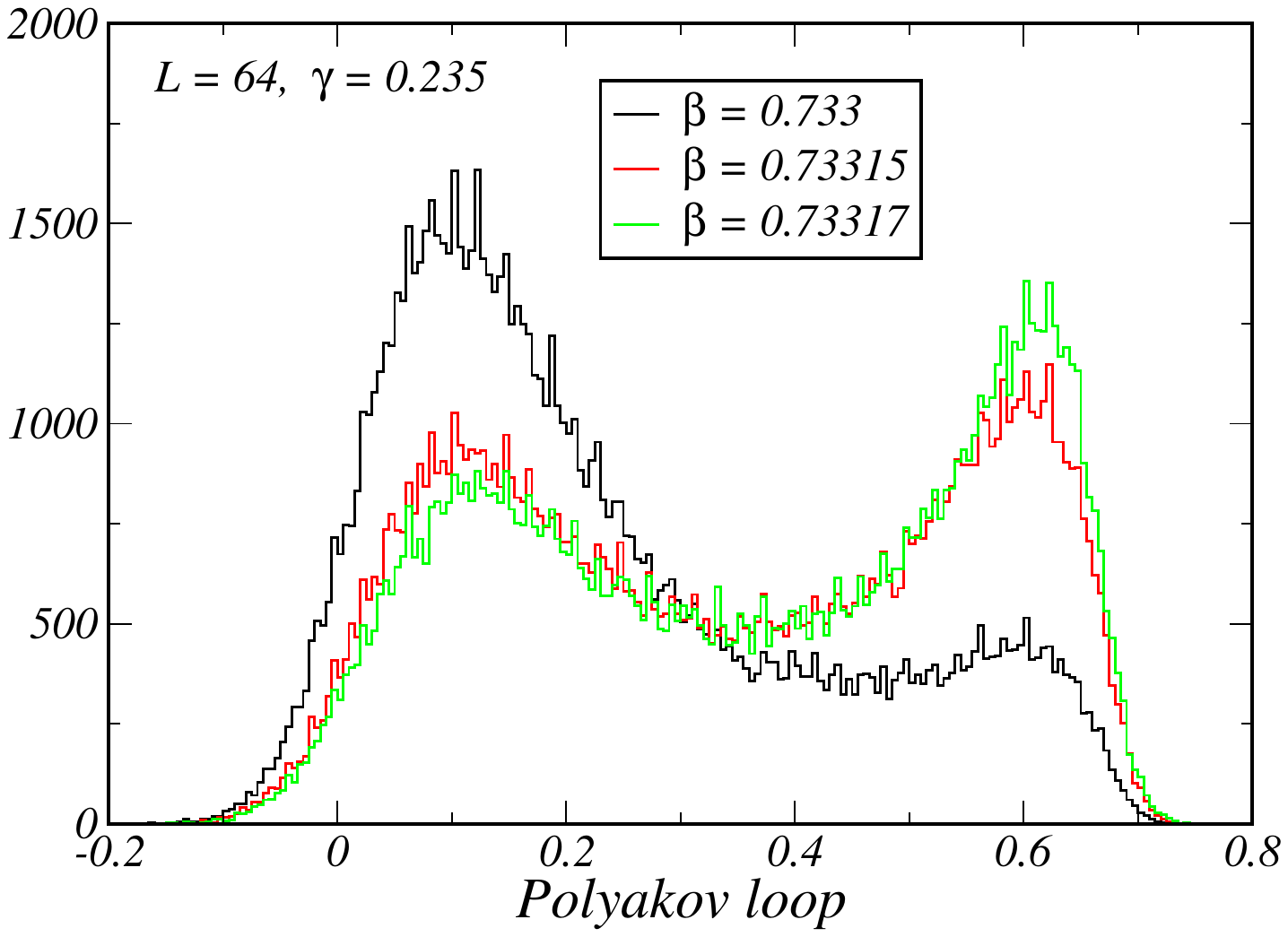}
\includegraphics[width=0.40\linewidth,clip]
{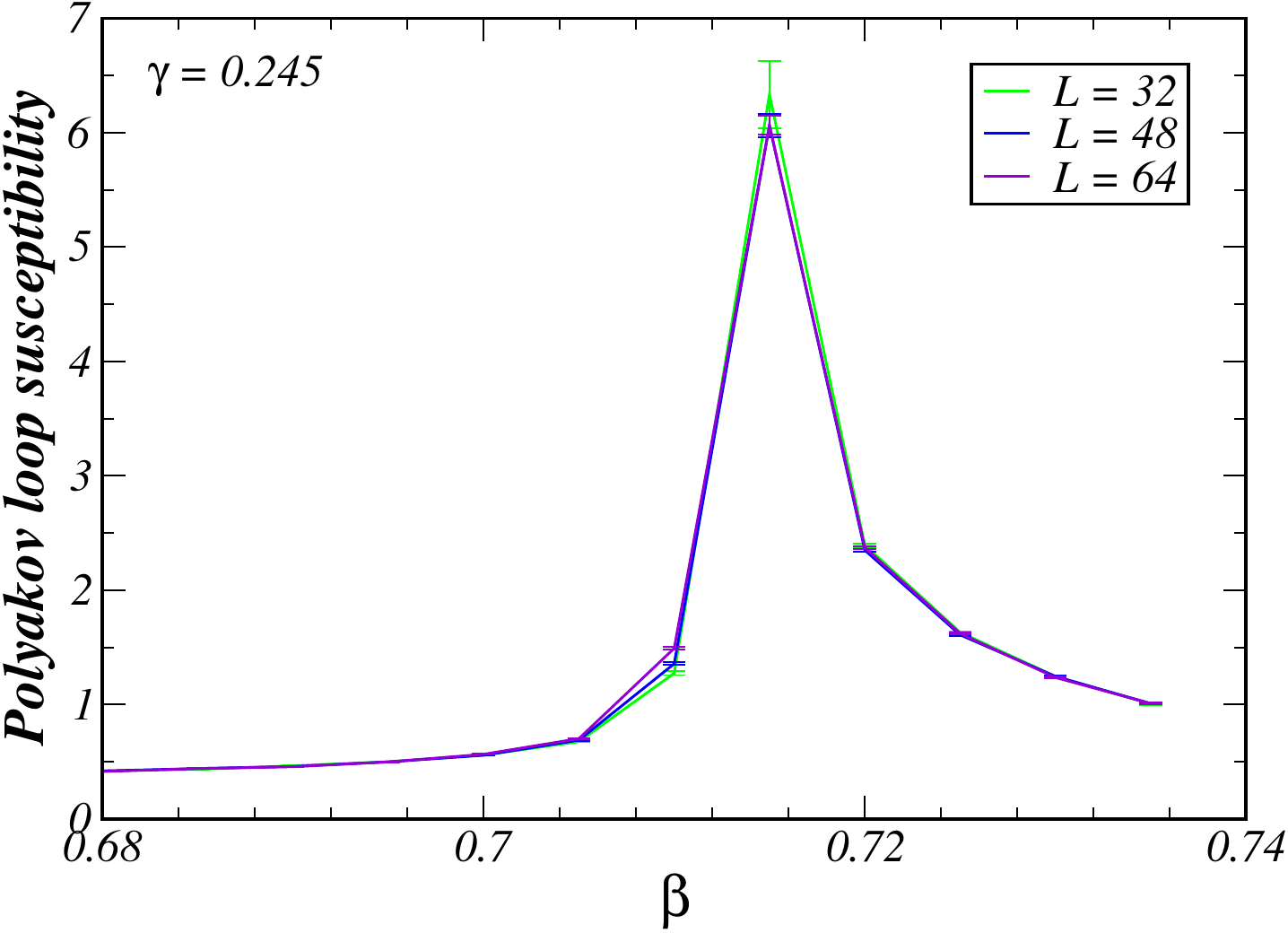}
\caption{Left: Histogram of the Polyakov loop in the region of the first order phase transition. Right: The Polyakov loop susceptibility in the crossover region. In both cases, the temporal extension of the lattice was $N_t=16$}.
\label{fig:pl_susc_crossover}
\end{figure}

\begin{figure}[H]
\centering
\includegraphics[width=0.8\linewidth,clip]{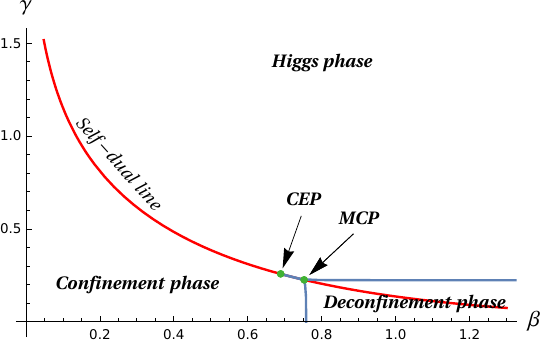}
\caption{Phase diagram of the $Z(2)$ gauge-Higgs theory on the lattice at finite temperature for $N_t=16$. The acronym CEP stands for ``critical end point" and MCP for ``multicritical point". For other details see text.}
\label{fig:phase_diagram_general}
\end{figure}

The left panel of Fig.\ref{fig:pl_susc_crossover} shows the Polyakov loop histogram 
for $\gamma=0.235$ for $N_t=16$. The two-peak structure indicates a first order transition.  Finally, the right plot of Fig.\ref{fig:pl_susc_crossover} presents the susceptibility in the crossover region, for $\gamma  > 0.240$. 
One observes that the peak of the susceptibility does not grow with the lattice volume. The resulting phase diagram for $N_t=16$ is shown in Fig.\ref{fig:phase_diagram_general}.

{\it The FM operator}. We have computed the FM operator on lattices with temporal extent $N_t=16$ and spatial sizes $L=32,48,64,96$. The number of measurements varied between 100k and 1200k per parameter set. Error analysis was performed by the jackknife method combined with blocking to account for data autocorrelation.
To study the FM operator across the deconfinement-Higgs phase transition we simulated the model at $\beta=1.0$ and several $\gamma$ values close to the critical point $\gamma_c=0.2235$. The behavior of the FM operator in the vicinity of the confinement-deconfinement transition was studied at $\gamma=0.21$ and several gauge couplings close to the critical point $\beta_c=0.7536$. In all plots below we use the notations $H_S$ and $H_T$ respectively for the spatial and temporal FM operators. For the temporal operator the default choice in our simulations was $T=N_t/2=8$. We checked that $H_T$ does not depend in a sizable way on other choices of $T$, such as $T=4$ and 6.

\begin{figure}[H]
\centering 
\includegraphics[width=0.48\linewidth,clip]
{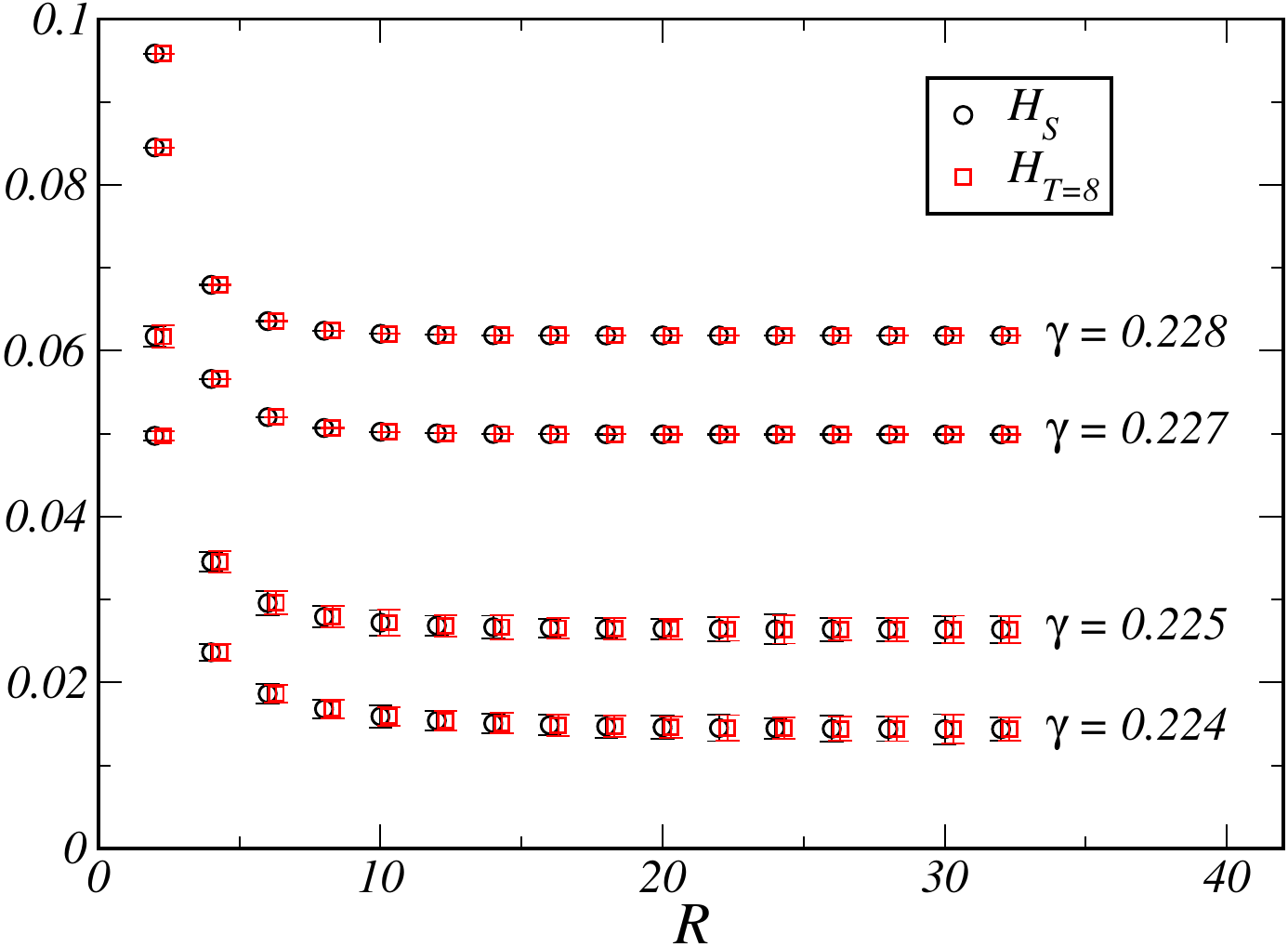}
\includegraphics[width=0.48\linewidth,clip]
{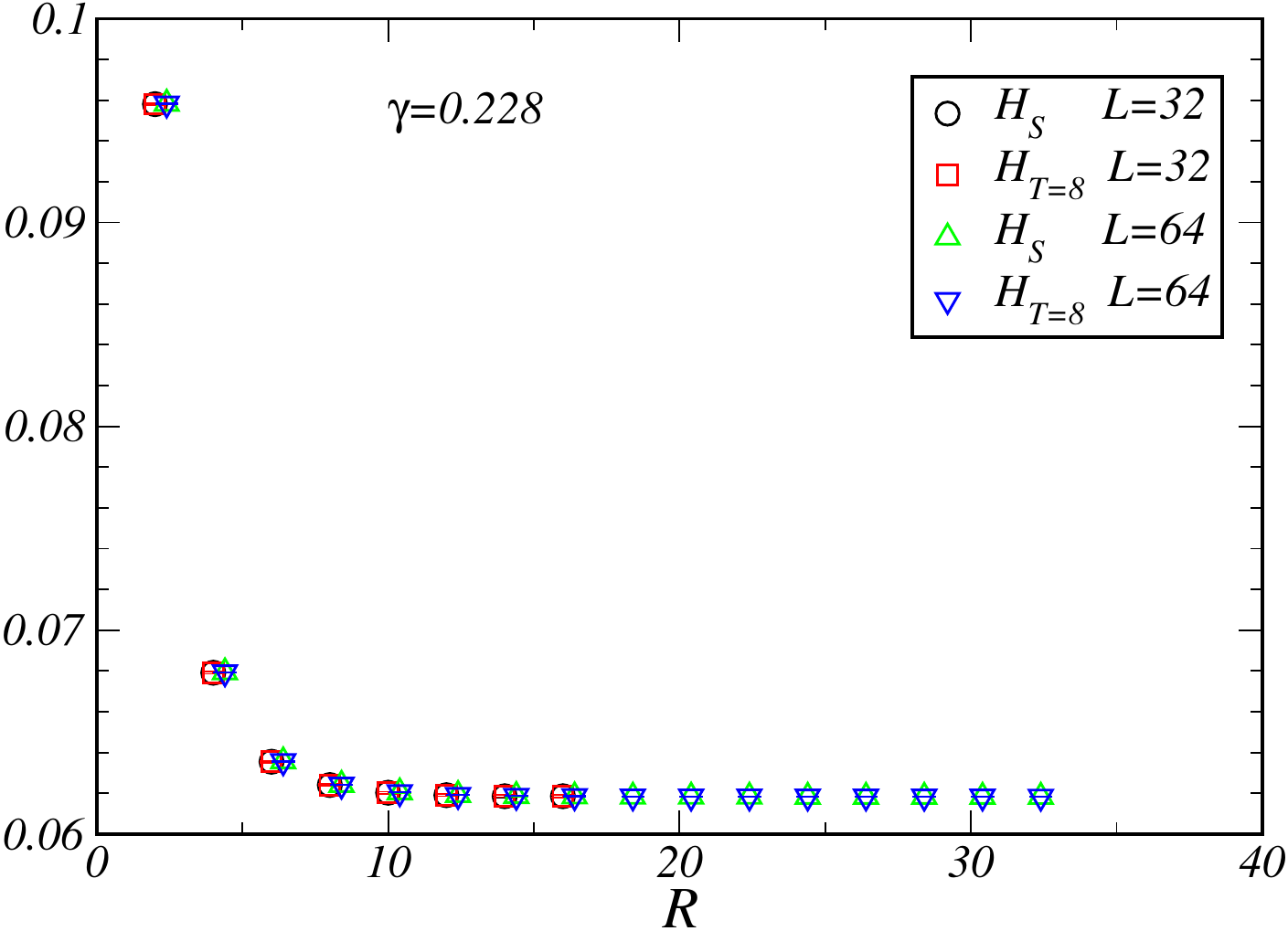}
\caption{Left panel: the FM operators in the Higgs phase above the deconfinement transition for $\beta=1$ and several $\gamma$ values on a spatial size $L=64$. Data for $H_T$ has been slightly shifted with respect to data for $H_S$ for clarity purposes. Right panel: the FM operator for $\beta=1$ and $\gamma=0.228$ on two spatial sizes. Data for $L=64$ slightly shifted horizontally with respect to data for $L=32$ for clarity purposes.}
\label{fig:FM_Higgs_deconf}
\end{figure}

\begin{figure}[H]
\centering
\includegraphics[width=0.32\linewidth,clip]
{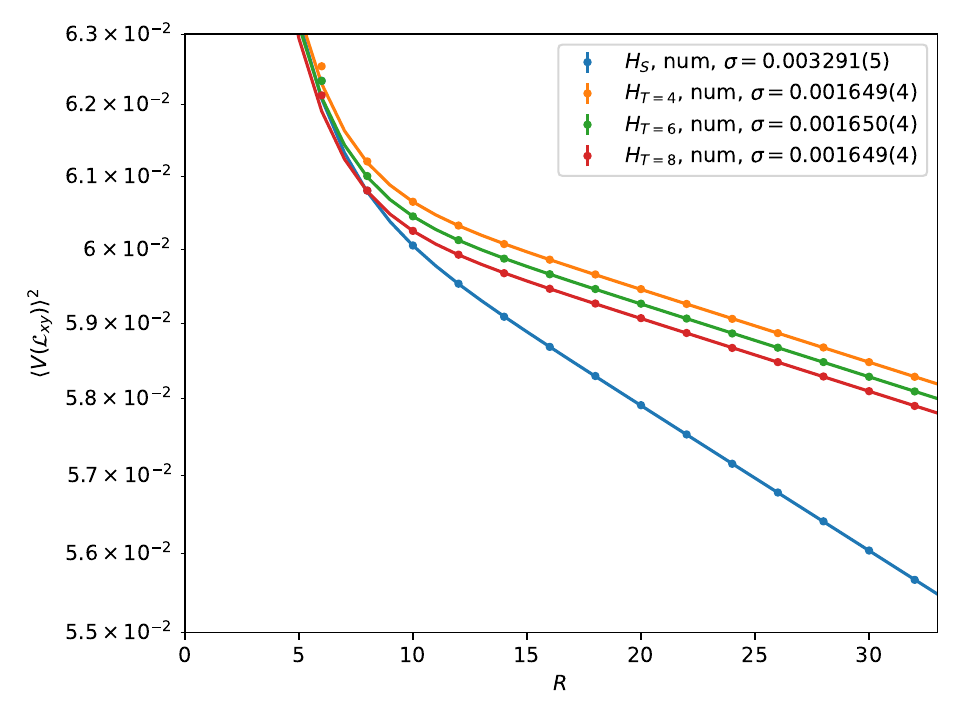}
\includegraphics[width=0.32\linewidth,clip]
{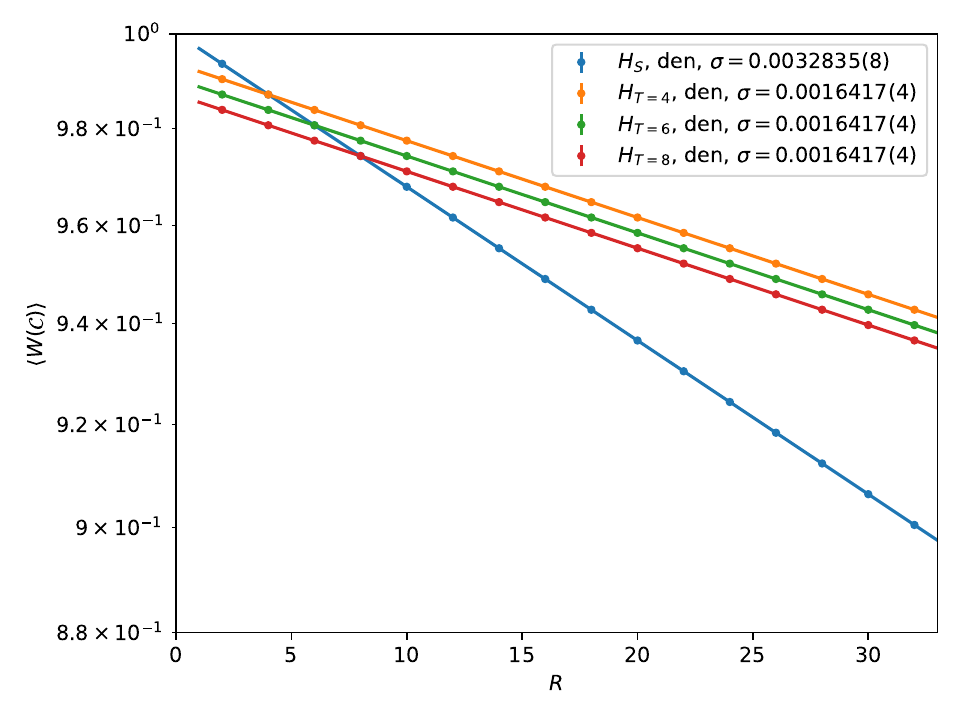}
\includegraphics[width=0.32\linewidth,clip]
{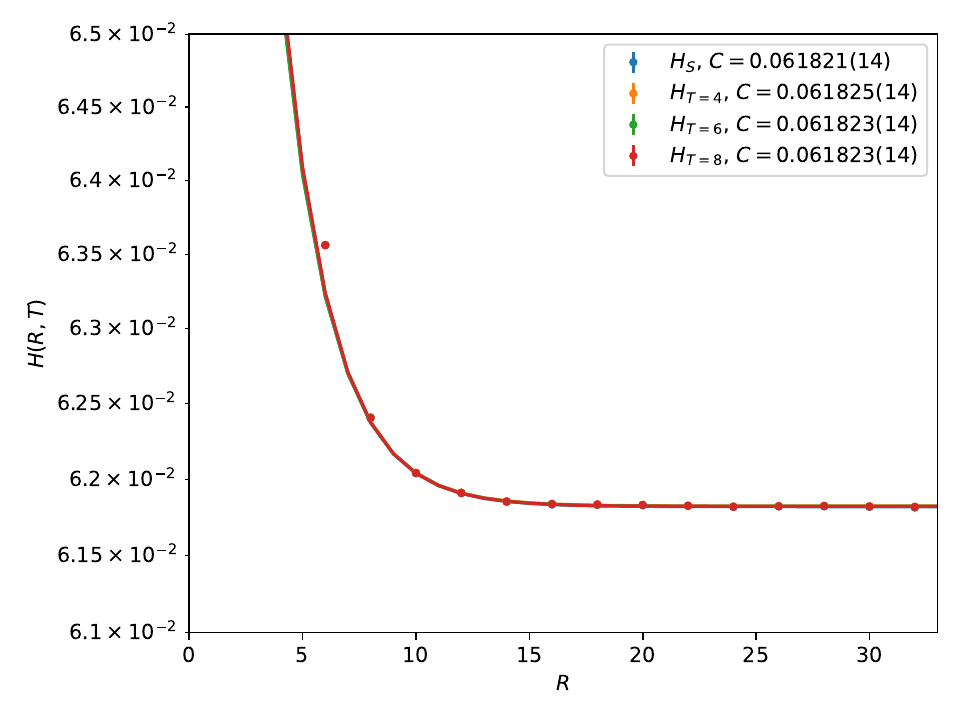}
\caption{
Fits of the numerator (left panel), denominator (middle panel) and the whole FM operator (right panel) to 
$\langle V({\cal{L}}_{xy}) \rangle^2 = A e^{-\sigma R} + B e^{-\sigma_2 R}$, $
\langle W({\cal{C}}) \rangle = A e^{-\sigma R}$, and $H(R, T) = A e^{-\sigma R} + C$, 
for $\beta=1.0$, $\gamma = 0.228$ on a $16 \times 64^2$ lattice (Higgs phase). In all cases the fit range in $R$ 
was set to $10 \leq R \leq 32$.}
\label{fig:FM_Higgs_deconf_fits}
\end{figure}

\begin{figure}[H]
\centering
\includegraphics[width=0.48\linewidth,clip]
{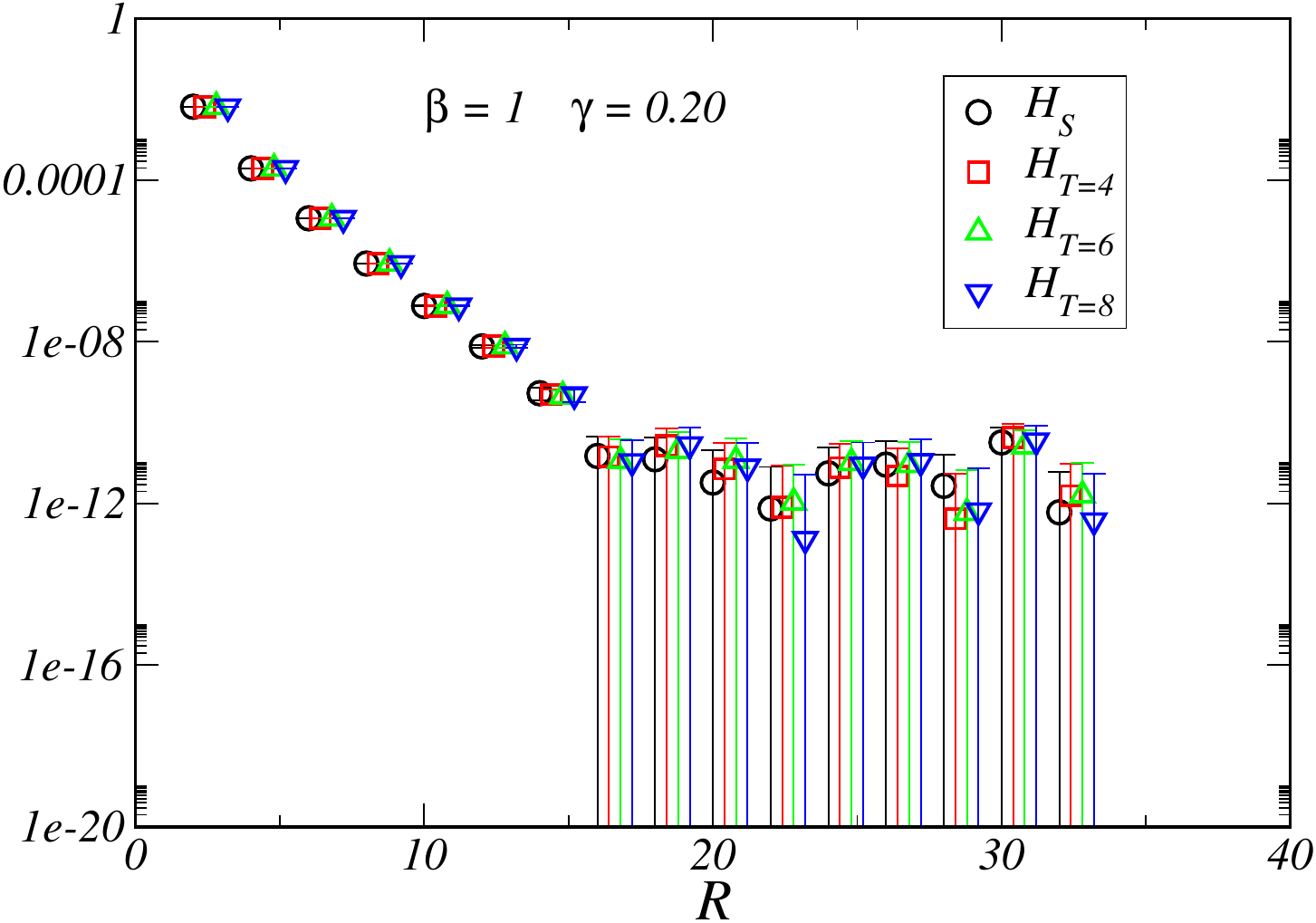}
\includegraphics[width=0.48\linewidth,clip]
{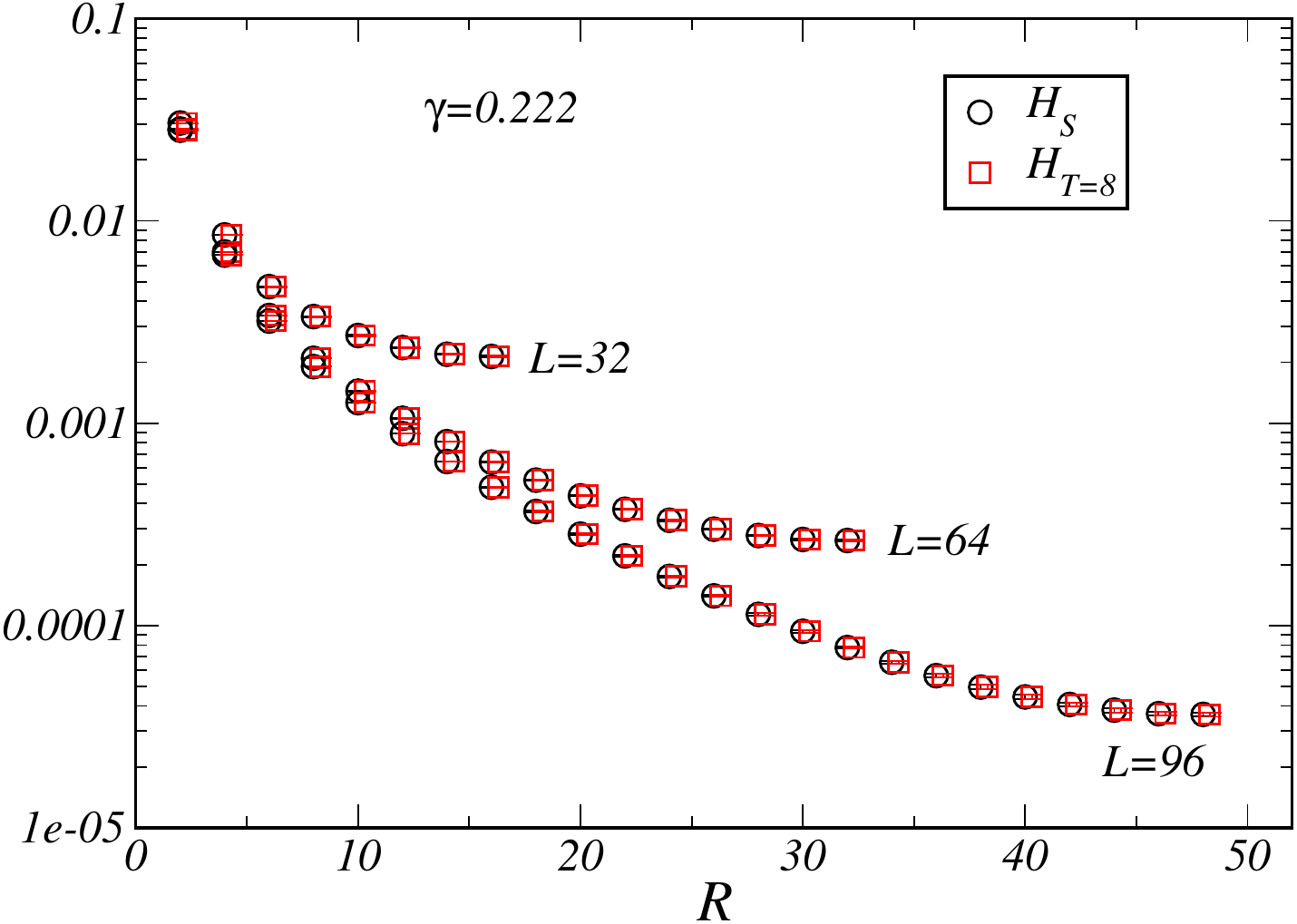}
\caption{Left panel: the FM operator deeply in the deconfined phase on the lattice for $\beta=1$ with the spatial size $L=64$. Right panel: the FM operator in the deconfined phase for $\beta=1$ close to the critical point on three spatial sizes. In both panels, data has been slightly shifted horizontally for clarity purposes, and a logarithmic scale has been used on the vertical axis.}
\label{fig:FM_deconf_Higgs}
\end{figure}

\begin{figure}[H]
\centering
\includegraphics[width=0.32\linewidth,clip]
{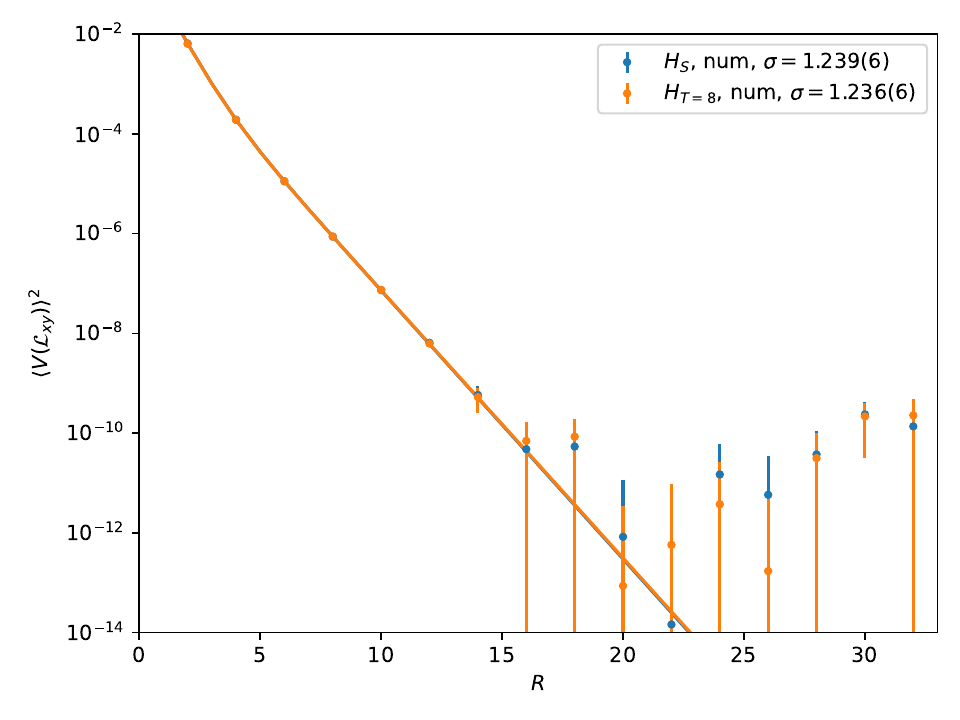}
\includegraphics[width=0.32\linewidth,clip]
{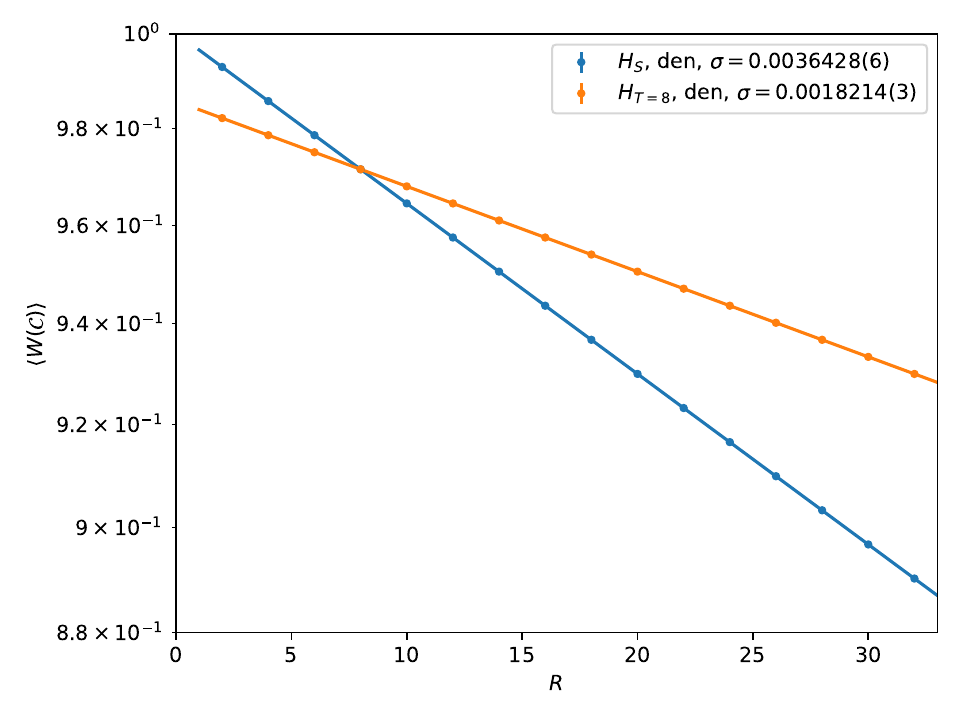}
\includegraphics[width=0.32\linewidth,clip]
{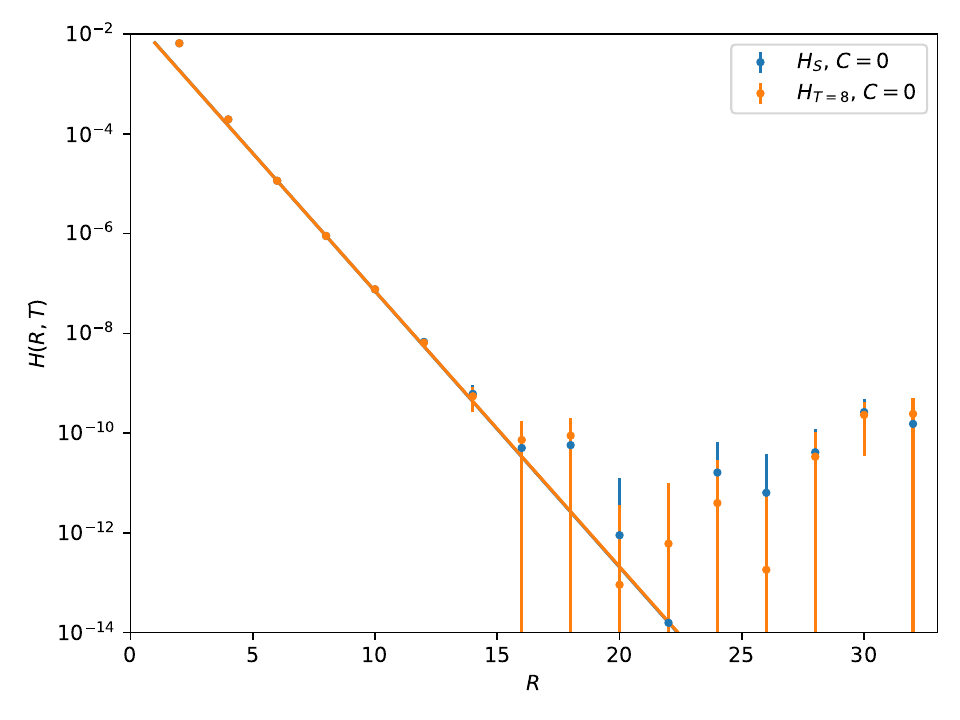}
\caption{
Fits of the numerator (left panel), denominator (middle panel) and the whole FM operator (right panel) to 
$\langle V({\cal{L}}_{xy}) \rangle^2 = A e^{-\sigma R} + B e^{-\sigma_2 R}$, $
\langle W({\cal{C}}) \rangle = A e^{-\sigma R}$, and $H(R, T) = A e^{-\sigma R}$, 
for $\beta=1.0$, $\gamma = 0.2$ on a $16 \times 64^2$ lattice (deconfinement phase). The fit range was taken to be 
$2 \leq R \leq 14$ (left), $6 \leq R \leq 32$ (middle), $6 \leq R \leq 14$ (right).}
\label{fig:FM_deconf_Higgs_fits}
\end{figure}

\begin{figure}[H]
\centering
\includegraphics[width=0.48\linewidth,clip]
{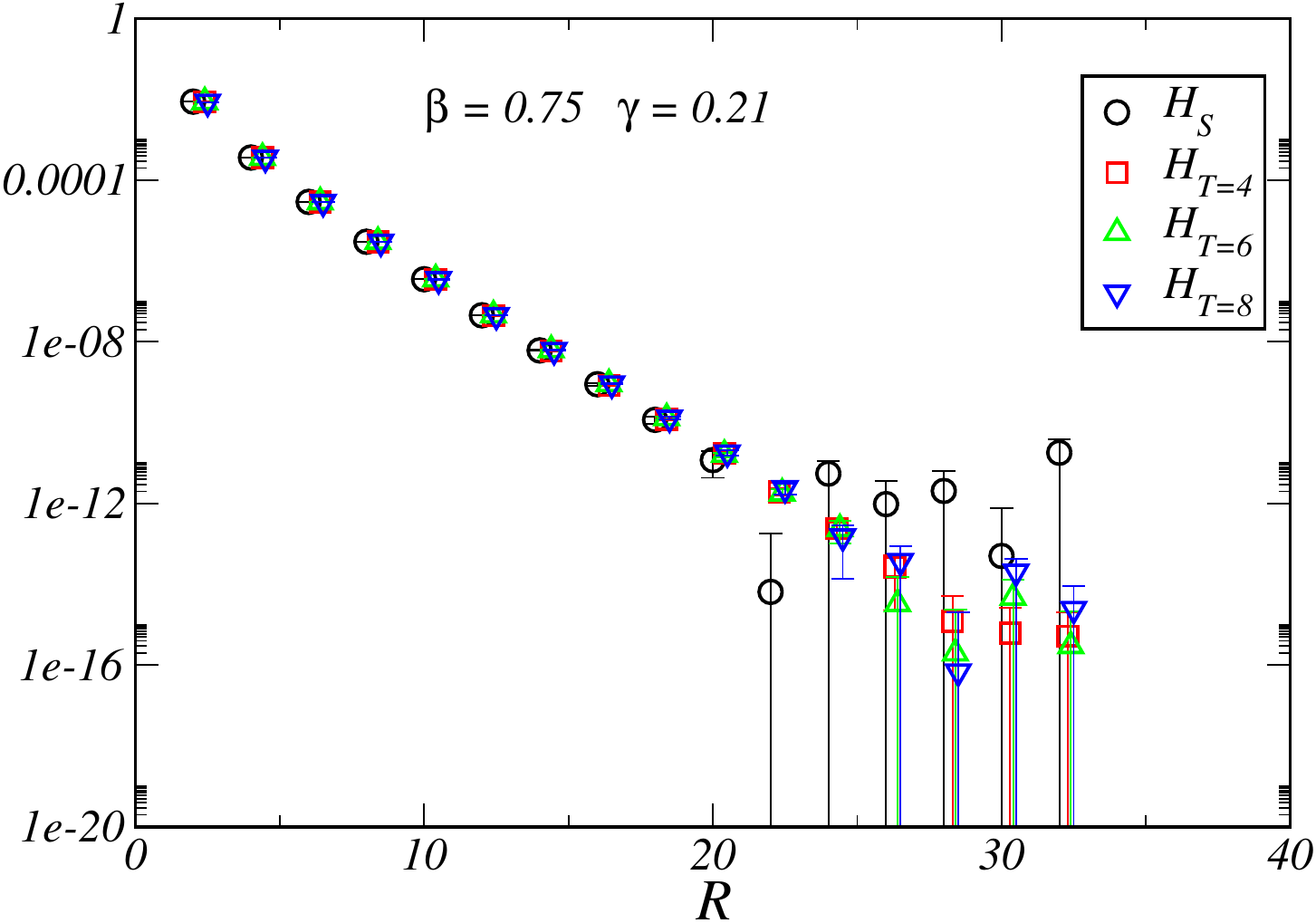}
\includegraphics[width=0.48\linewidth,clip]
{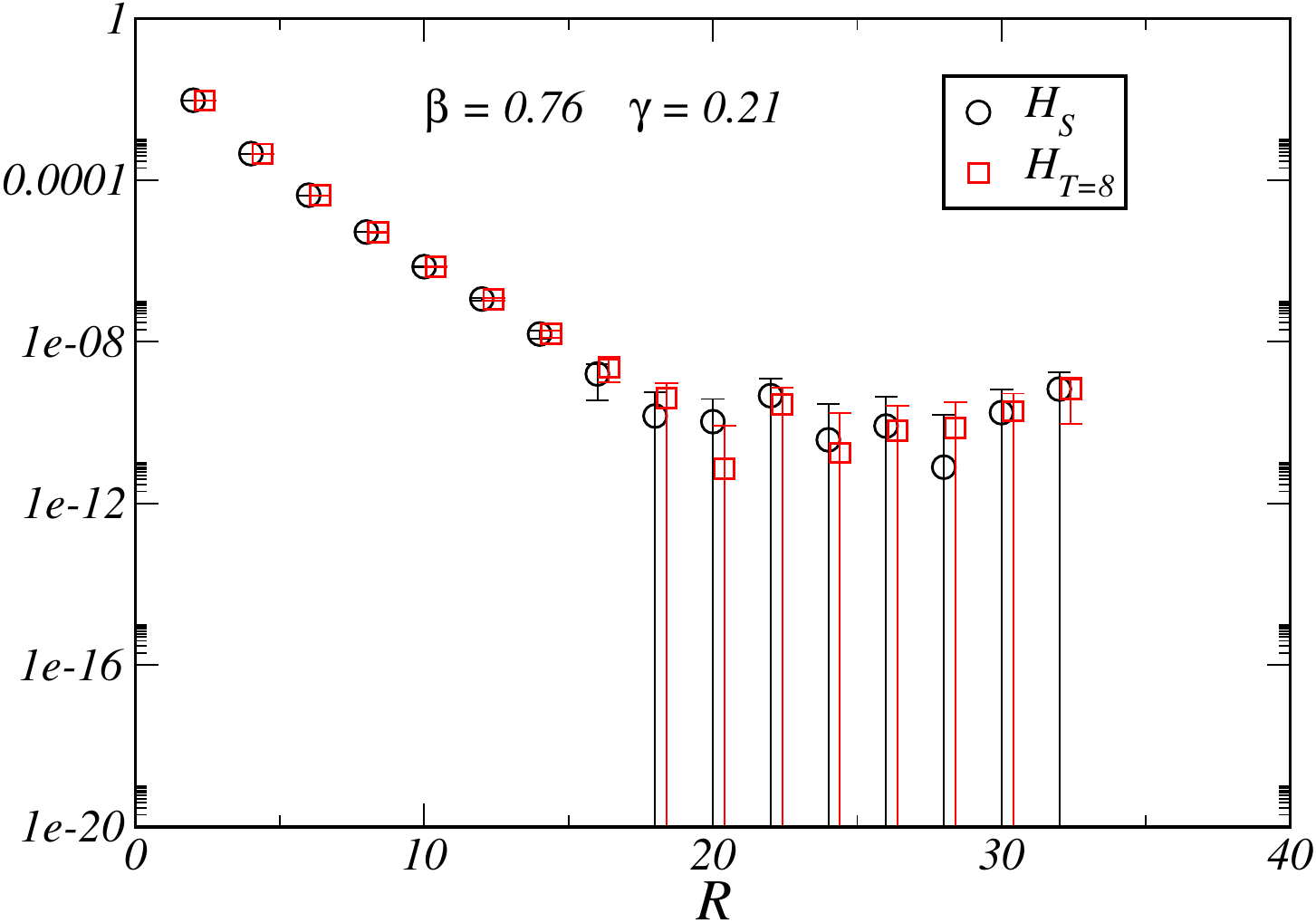}
\caption{The FM operators in the confinement (left panel) and in the deconfinement (right panel) phase near the confinement to deconfinement transition at $\gamma=0.21$ on $L=64$ lattice. The critical point is at $\beta=0.75314$. In both panels data has been slightly shifted horizontally for clarity, and a logarithmic scale has been used on the vertical axis.}
\label{fig:FM_conf_deconf}
\end{figure}

\begin{figure}[H]
\centering
\includegraphics[width=0.32\linewidth,clip]
{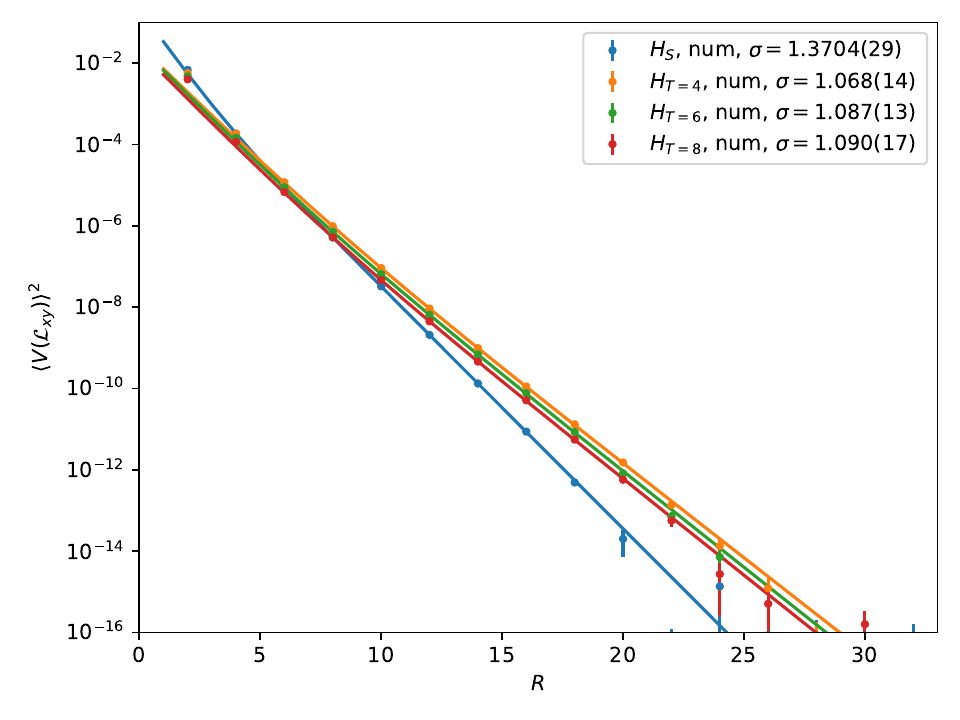}
\includegraphics[width=0.32\linewidth,clip]
{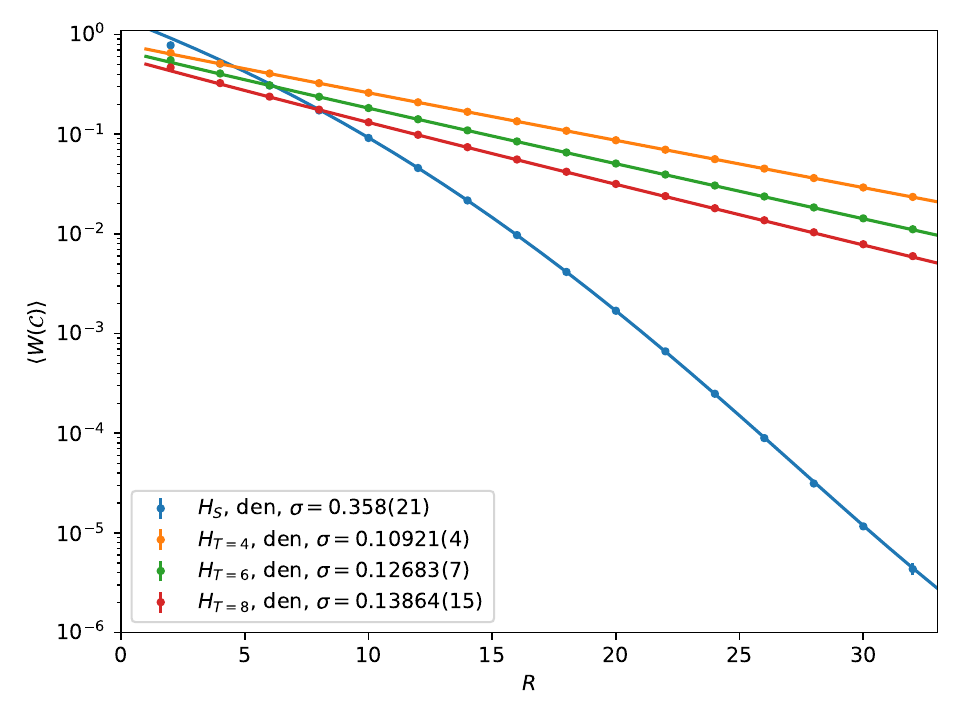}
\includegraphics[width=0.32\linewidth,clip]
{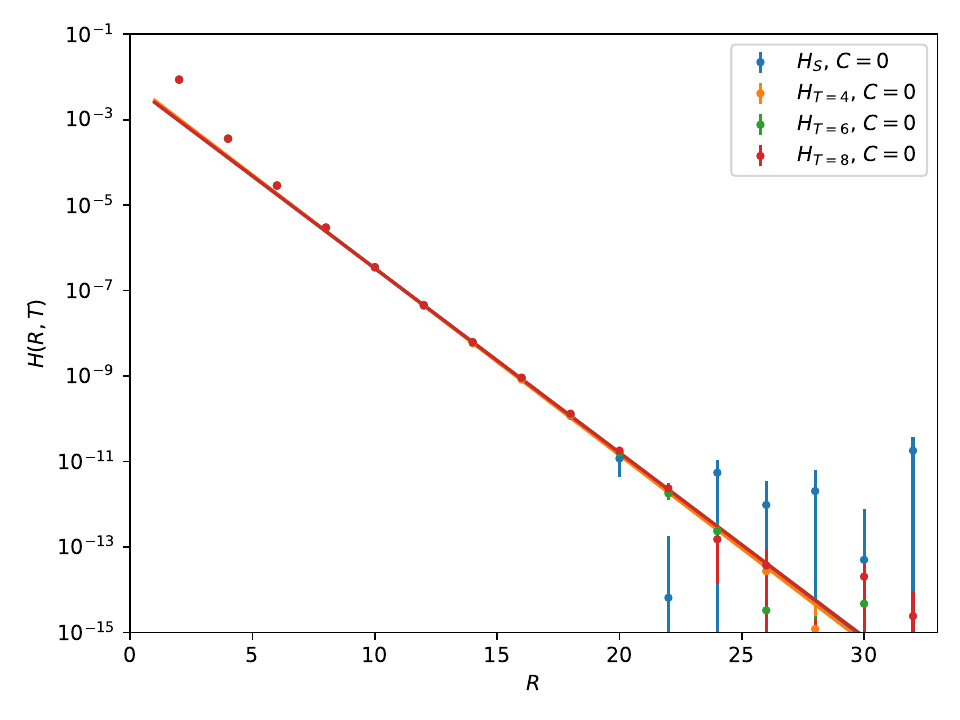}
\caption{
Fits of the numerator (left panel), denominator (middle panel) and the whole FM operator (right panel) to 
$\langle V({\cal{L}}_{xy}) \rangle^2 = A e^{-\sigma R} + B e^{-\sigma_2 R}$, $
\langle W({\cal{C}}) \rangle = A e^{-\sigma R} + B e^{-\sigma_2 R T - \alpha R}$, and $H(R, T) = A e^{-\sigma R}$, 
for $\beta=0.75$, $\gamma = 0.21$ on a $16 \times 64^2$ lattice (confinement phase). The fit range was taken to be 
$6 \leq R \leq 16$ (left), $10 \leq R \leq 32$ (middle), $8 \leq R \leq 16$ (right).}
\label{fig:FM_conf_deconf_fit}
\end{figure}

\begin{figure}[H]
\centering
\includegraphics[width=0.32\linewidth,clip]
{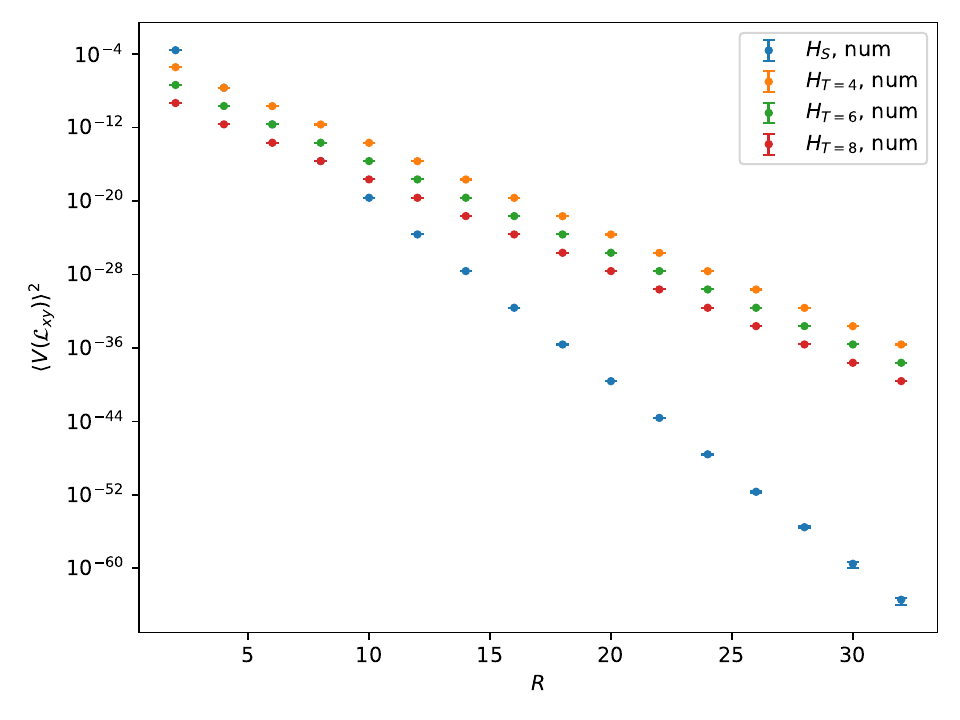}
\includegraphics[width=0.32\linewidth,clip]
{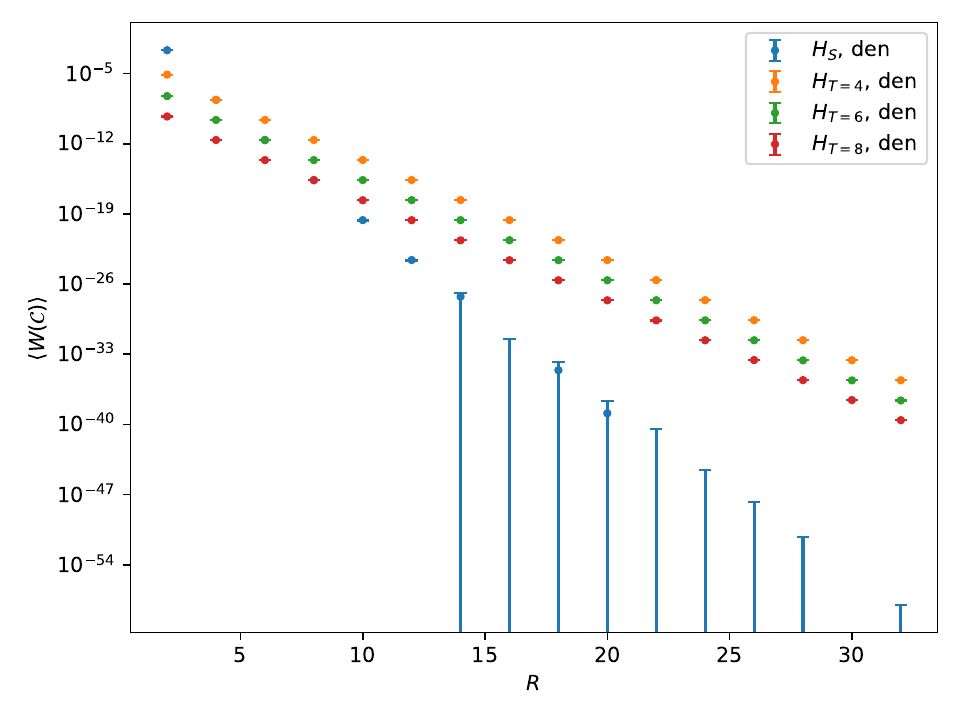}
\includegraphics[width=0.32\linewidth,clip]
{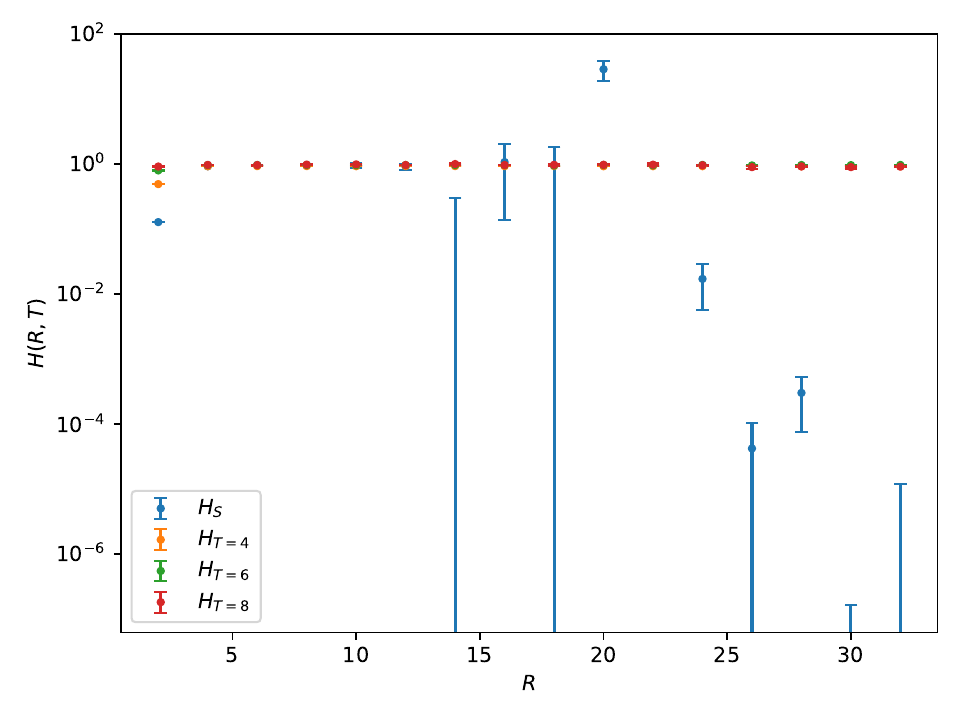} \\
\includegraphics[width=0.32\linewidth,clip]
{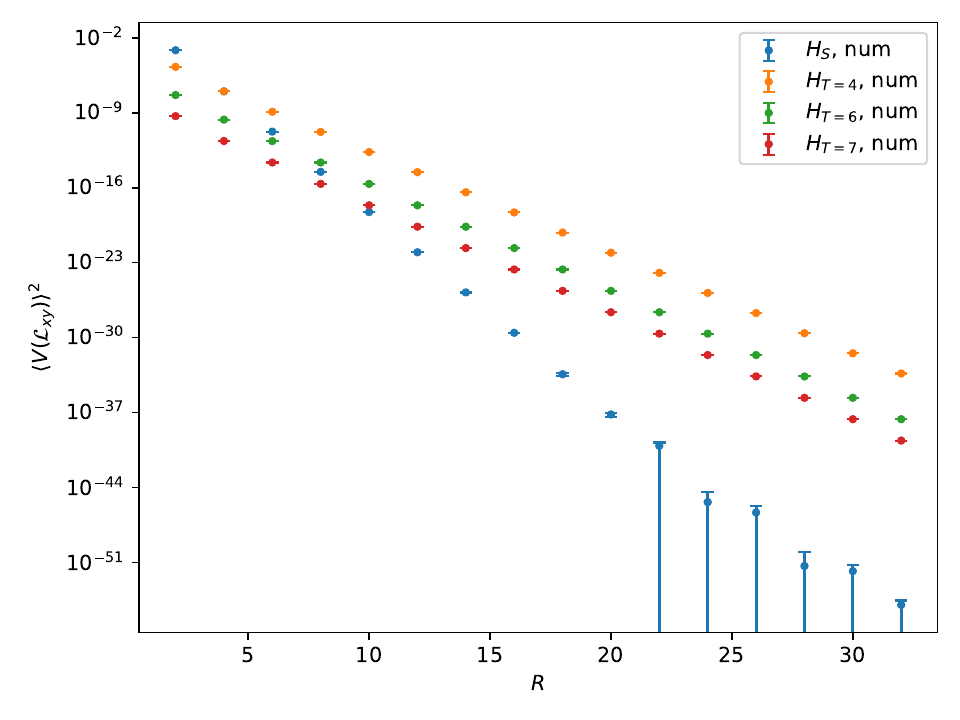}
\includegraphics[width=0.32\linewidth,clip]
{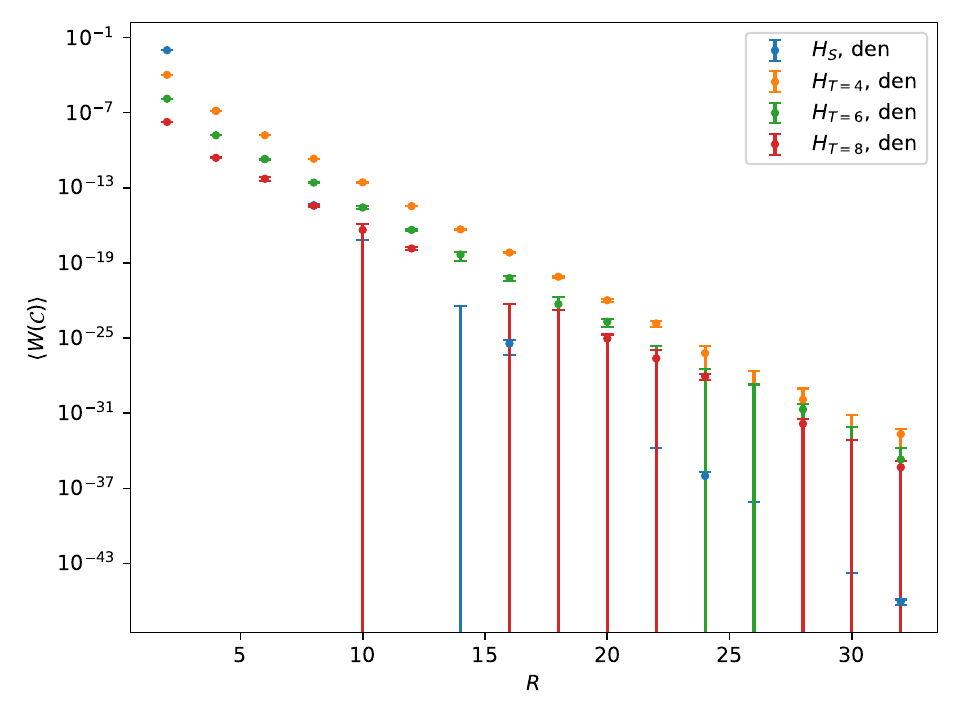}
\includegraphics[width=0.32\linewidth,clip]
{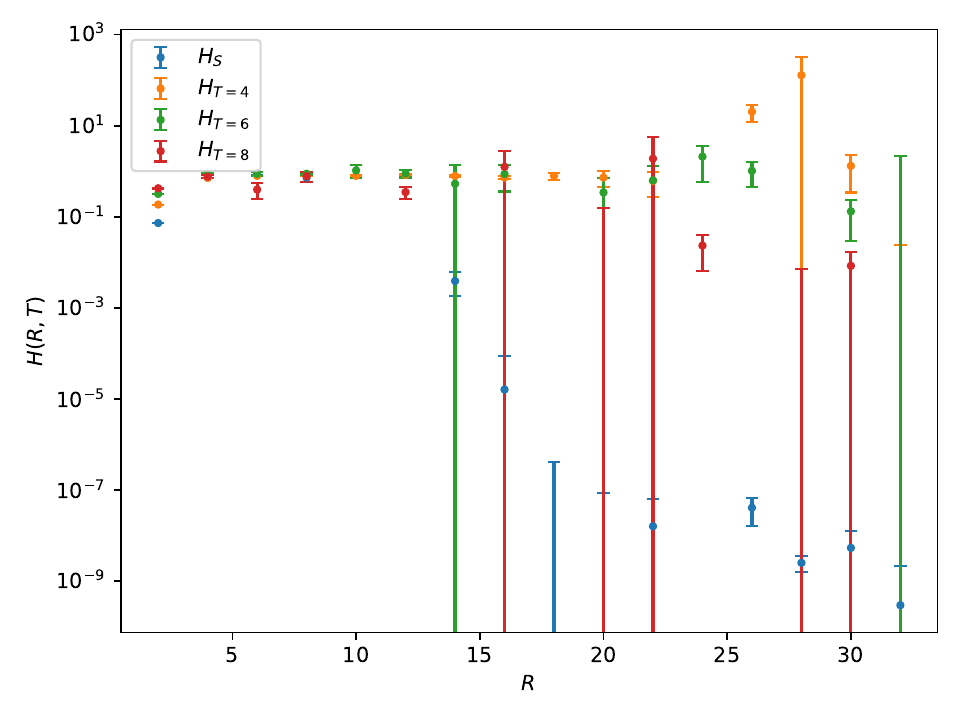} \\
\includegraphics[width=0.32\linewidth,clip]
{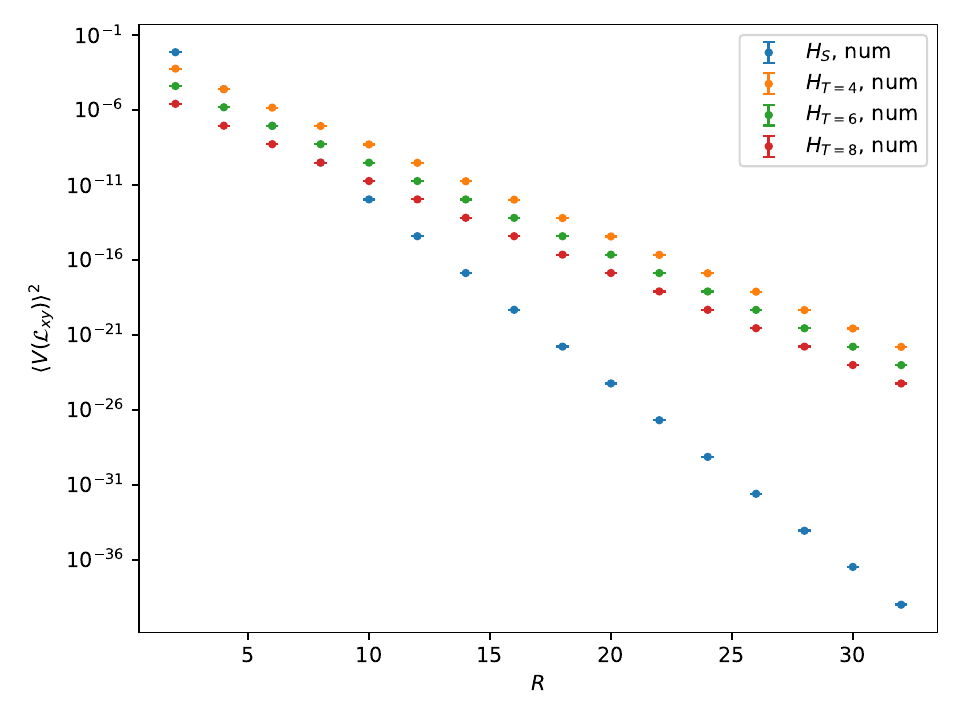}
\includegraphics[width=0.32\linewidth,clip]
{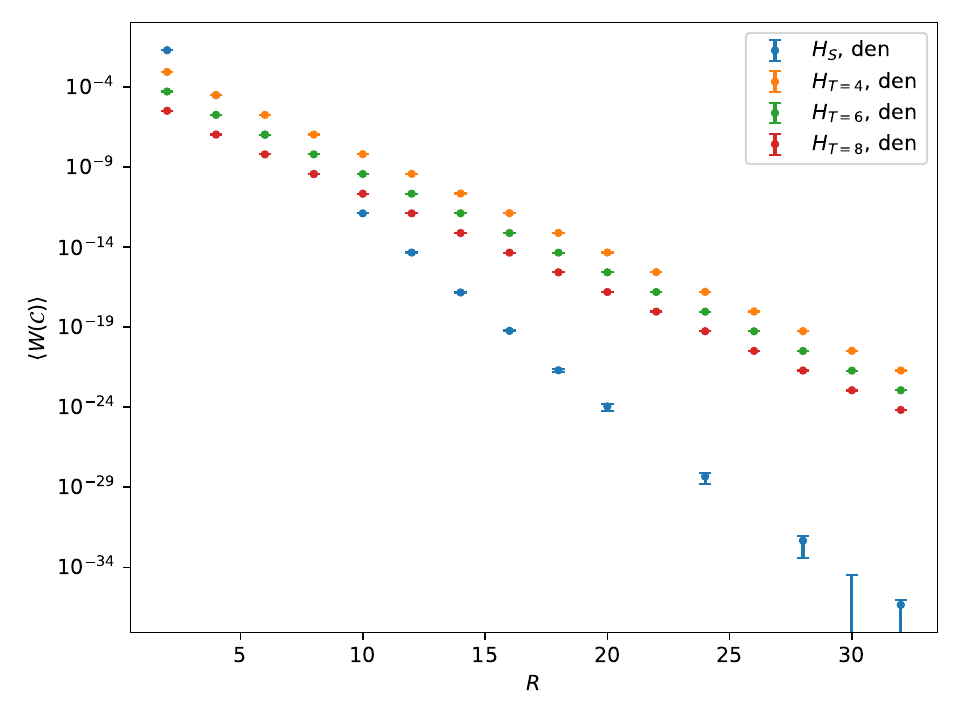}
\includegraphics[width=0.32\linewidth,clip]
{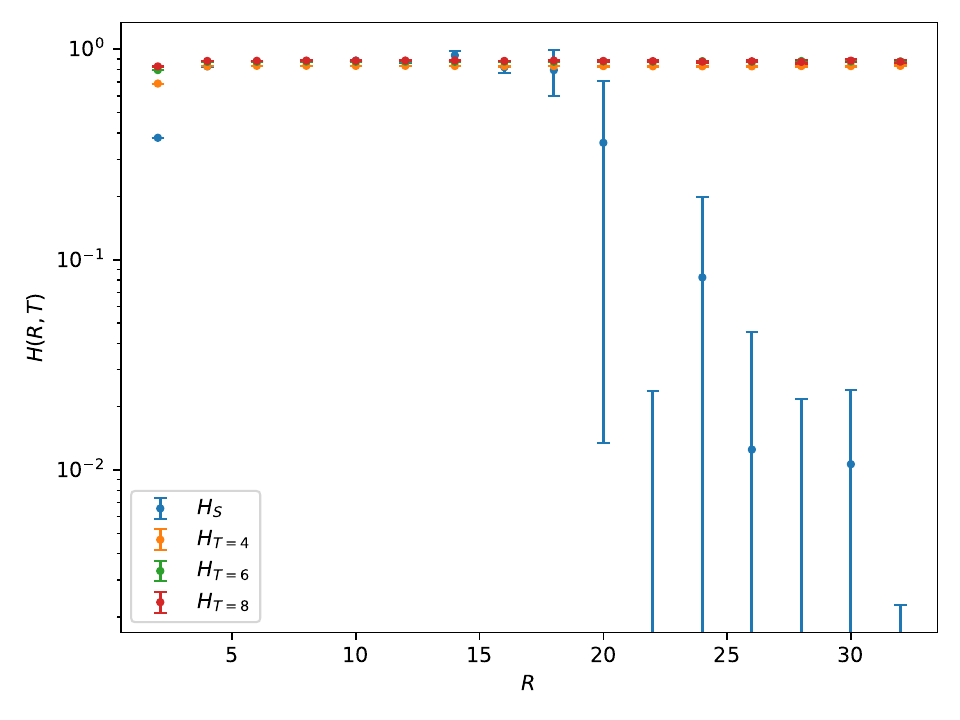}
\caption{
Numerators (left column), denominators (middle column) and the whole FM operators (right column)
deep in the confinement region: $\beta=0.2$, $\gamma=0.3$ (top row), 
$\beta=0.3$, $\gamma=0.3$ (middle row),
$\beta=0.3$, $\gamma=0.4$ (bottom row).
}
\label{fig:FM_deep_conf_deconf}
\end{figure}

 To decrease the variance of the observables we used a ``multihit'' method~\cite{PPR_multihit}. It consists in replacing each link variable in the operator (or the pair of link variables forming each of the corners) with their averages taken by keeping fixed the link variables which share the same plaquette with them. To decrease the autocorrelation time and avoid the simulation trajectory getting ``stuck'' around a local energy minimum, a cluster update~\cite{SW_cluster} for the Higgs fields was employed. 
 The cluster update for the Higgs field allows us to build a ``cluster improved'' Wilson line observable, following~\cite{cluster_improved_observables}, by analytically averaging over all independent cluster flips, noticing that contributions from Wilson lines with ends belonging to the same cluster remain unchanged, while those with ends belonging to different clusters vanish exactly. Since the ``cluster improved'' observable is incompatible with the ``multihit'' method, we used the ``cluster improvement'' only for $\beta=0.75$ and $\gamma=0.21$, where it allows one to significantly improve the signal on larger distances.

The results of simulations are shown in Figs.~\ref{fig:FM_Higgs_deconf}, \ref{fig:FM_deconf_Higgs},  \ref{fig:FM_conf_deconf}. In all cases data are shown for $R$ even and from $R=2$ to $R=L/2$.
Data is shown for $T=8,6,4$. Statistics was about 1M.
The left panel of Fig.\ref{fig:FM_Higgs_deconf} presents the behavior of the FM operator in the Higgs phase above the critical point on a lattice with spatial size $L=64$. Clearly, both the spatial and the temporal operators tend to a constant with increasing distance. To verify this, we performed fits of numerator, denominator, and the whole expression (\ref{FMoperator_def}) to $\langle V({\cal{L}}_{xy}) \rangle^2 = A \exp{(-\sigma R)} + B \exp{(-\sigma_2 R)}$, $
\langle W({\cal{C}}) \rangle = A \exp{(-\sigma R)}$, and $H(R, T) = A \exp{(-\sigma R)} + C$. The results are shown in Fig.\ref{fig:FM_Higgs_deconf_fits}. One can see that the values of $\sigma$ for the numerator and denominator agree within two standard errors. Furthermore, the fit of the FM operator itself clearly shows a nonzero constant $C$. This constant increases with the value of the Higgs coupling $\gamma$, as expected. We have also checked whether there is a volume dependence of the constant. The typical behavior is shown on the right plot of Fig.\ref{fig:FM_Higgs_deconf}, which tells us that the volume dependence of the FM operator is negligible in the Higgs phase. 
In Fig.\ref{fig:FM_deconf_Higgs} we show the FM operator at the same gauge coupling $\beta=1.0$ for two $\gamma$ values in the deconfined phase. Deeply in this phase (left panel) the FM operator on the lattice $L=64$ decreases to a value that is compatible with zero (note the logarithmic scale). The right panel shows the FM operator close to the critical point. 
Both operators decrease slowly after some large distance and one could conclude that they approach a constant when $R$ is large enough. However, here we have found an essential volume dependence: this constant becomes smaller with increasing volume and might well vanish in the thermodynamic limit. We could not go to larger lattices to check this hypothesis further, but data in Fig.\ref{fig:FM_deconf_Higgs} strongly support this scenario. 
The fits for $\beta=1.0$, $\gamma=0.2$ are shown in Fig.\ref{fig:FM_deconf_Higgs_fits}. Here, due to the rapid decrease of the numerator, we had to restrict the fit range for the numerator (and consequently for the full FM operator) to $R \leq 14$, in order to prevent the covariance matrix of the data from becoming singular.
Here we see that the values of $\sigma$ for the numerator and denominator differ by more than 200 standard errors, which implies that $\lim_{R \to \infty} H(R, T) = 0$. The fits with free value of $C$ for $H(R, T)$ are extremely ill-behaved, so the right panel shows the fit with $C$ fixed to zero (which is supported both by the data itself, and by the separate numerator and denominator fits).
Finally, in Fig.\ref{fig:FM_conf_deconf} we have depicted the FM operators in the vicinity of the confinement-deconfinement phase transition for $\gamma=0.21$ and two values of the gauge coupling on a $L=64$ lattice. 
In both phases the FM operators tend to a small constant compatible with zero. This is expected in the deconfined phase. 

For the confinement phase, we again can analyze the behavior of numerator and denominator in the FM operator separately.
The fits are shown in Fig.\ref{fig:FM_conf_deconf_fit}. We again had to limit the fitting range from above both for the numerator fits, and for the full FM operator fits, due to the covariance matrix getting singular. For the Wilson loop behavior (denominator), it is easy to see that at small distances the spatial Wilson loop exhibits area-law behavior, proving that we indeed stay in the confinement regime and motivating the addition of a term $B \exp{(-\sigma_2 R T - \alpha R)}$ to the fitting function. This term should be irrelevant in the large-$R$ regime, since the perimeter-law term $A \exp(-\sigma R)$ will dominate it at $R$ large enough, causing string breaking at $R > R_{\rm sb} \sim \sigma / \sigma_2$. In our case 
$\sigma / \sigma_2 \approx 50$, so the string breaking effect is not visible at available distances, hiding the true value of the perimeter-law constant $\sigma$.  We also point out the fact that attempts to introduce an area-law behavior to the 
numerator fitting function do not result in good fits to our data -- either because the Wilson line does not exhibit a sizable area-law behavior, or because the range of $R$ where we can reliably estimate the numerator is too narrow to constrain all the fit parameters. 
The fit results show that the values 
of $\sigma$ for the numerator are significantly higher than the ones for the denominator, forcing us to conclude that 
$\lim_{T \to \infty} H(R, T) = 0$, despite the system being in the confinement phase. Our interpretation of this result is that the string-breaking distance in the confinement region is significantly higher than the distances studied here. Proving this scenario would require a significant increase in both the precision of the FM operator estimation and in the lattice size. 
We reiterate that in the region of convergence of the strong coupling expansion, at small values of $\beta$, the FM operator goes to a constant. We have also performed simulations deeply in the confined phase for $\beta=0.1, 0.3$. For all $\gamma > \beta$ both FM operators approach 1 with distance very fast ({\it e.g.} $H(R,R)=0.9982(16)$ at $R=16$ for $\beta=0.1$ and $\gamma=1$, in agreement with analytic calculations), while for $\gamma < \beta$ they rapidly decrease to a very small value. The presented results also reveal that $H_T$ depends very weakly on $T$.

To better investigate the behavior of the FM operators in the confinement phase we performed a number of simulations at small values of $\beta$ and $\gamma$ -- far from the deconfinement boundary. 
In this case both expectation values in the definition of the FM operators become exponentially small with the Wilson loop size, 
making their direct determination prohibitively expensive, even with the improvements described above, 
as statistical fluctuations start dominating the signal already at relatively small values of $R$. 
To improve the precision of the measurements, we implemented a variant of the L\"uscher-Weisz multilevel algorithm~\cite{multilevel}.
The lattice was decomposed into spatial slices of width 4. Links surrounded by other links in the slice will be called ``inner links'' and their updates
``inner updates'', while links lying on the boundary will be called ``outer links'' and their updates ``outer updates''. During the inner updates the values of the fields on the slice 
boundaries were kept fixed, while the fields inside each slice were updated independently. 
For each outer configuration we generated $n_{\mathrm{inner}} = 1000$ statistically independent measurements for each slice,
and 
replaced the contribution to the Wilson loop and Wilson line from each slice by its conditional expectation value for a given 
boundary field configuration. The full observable is then reconstructed as the product of the averaged contribution from all
slices. To avoid the increase of variance coming from the Wilson loops (or Wilson lines), that contain outer links, only the observables that do not contain such links were included in the volume average. 
This procedure was repeated for $n_{\mathrm{outer}} = 800$ outer configurations, separated by 40 global updates of the whole lattice (outer updates). 
This multilevel estimator suppresses statistical fluctuations associated with large Wilson loops, allowing us to reliably measure FM operators deep in the confinement phase for much larger $R$ than it was possible with our previous approach. The resulting measurements are shown in Fig.\ref{fig:FM_deep_conf_deconf}. 
Deep in the confinement phase, both the spatial and temporal FM ratios approach a nonzero constant as $R$ grows.

The efficiency of the multilevel algorithm deteriorates as $\gamma$ decreases. For example, at $\beta=0.3$ and $\gamma=0.2$ the signal-to-noise ratio is still insufficient to obtain reliable results beyond $R=4$. Thus, our simulations do  not allow us to determine 
whether the FM operator remains nonzero for such small $\gamma$.

\section{Summary} 

In this paper we studied the $Z(2)$ gauge-Higgs lattice model at finite temperature. Our main goal was to clarify if the Fredenhagen-Marcu operator can be used as a candidate order parameter at finite temperature to distinguish the deconfined phase from the rest of the phase diagram. The crucial point is to prove that the large-distance limit of the FM operator vanishes in the deconfined phase, while it tends to a constant in the confinement and Higgs phases.

A similar analysis has been done by some of us for the $(3+1)$-dimensional $SU(2)$ gauge-Higgs LGT in a recent paper~\cite{su2_fm_fin_temp} where, moreover, a simple non-rigorous argument was advanced to explain why the large distance limit of the FM operator can vanish in the deconfined phase, and its behavior was studied by numerical simulations. The results obtained were, however, not conclusive as larger lattices are needed to get reliable predictions in this model. Therefore, we decided to study the simpler $Z(2)$ LGT, where larger lattice sizes and statistics can be employed  without much effort.  Our main findings can be briefly summarized as follows. 

\begin{itemize}
\item 
We have studied the phase diagram of the $Z(2)$ gauge-Higgs theory by employing local observables like the Polyakov loop, expectation values of the plaquette and Higgs action. This allowed us to locate the critical lines of phase transitions and check the universality class of the model. The collapse plots for the susceptibility of the Polyakov loop show that the model is in the universality class of $2d$ Ising model. This ensures that the spatial lattice volume is large enough, so the system can indeed be considered as a finite-temperature system.   
\item 
The large-distance behavior of both the spatial and temporal FM operators in the deconfined phase seems to be consistent with zero, especially deeply in this phase, {\it i.e.} at small values of the Higgs coupling. Close to the critical point the FM operators decrease rather slowly with the distance and reach a constant value for the largest distance away from zero. However, this constant decreases with the volume and seems to tend to zero. 
\item 
Our data suggest that both the spatial and temporal FM operators are compatible with zero also in the confinement phase, close to the deconfinement phase transition.
In contrast, simulations performed deep in the confined phase show that the FM operators tend to a nonzero constant value with increasing $R$, 
at least for the region $\gamma \geq \beta$. It remains unclear whether there is a line inside the confined phase 
that separates the region with vanishing and nonvanishing FM operators, 
or whether the FM operator remains nonzero throughout the confined phase, but becomes too small to resolve in our simulations near the deconfinement transition. The existence of two subregions within the confined phase, separated by a crossover or transition line, would carry nontrivial conceptual consequences and cannot currently be ruled out based on our data.

\end{itemize}

The deconfinement phase transition in QCD is associated with a crossover transition at physical quark masses. As we already emphasized, this cannot rule out a non-analytic behavior of some non-local observables. Therefore,  
as a next step in our investigation of the properties of FM operators, we intend to study its behavior in gauge models coupled with fermions. This can potentially give a direct access to the detection of the deconfinement phase in QCD at finite temperature. 

\vspace*{1cm}

{\bf Acknowledgments}. We thank the anonymous referee for constructive comments and suggestions that improved the manuscript. Authors BA, OB and AP acknowledge support from the INFN/NPQCD project.
This work is partially supported by ICSC -- Centro Nazionale di Ricerca in High Performance Computing, Big Data and Quantum Computing, funded by European Union -- NextGenerationEU. 
Most numerical simulations have been performed on the CSN4 cluster of the Scientific Computing Center at INFN-Pisa. 
VC acknowledges support by the Deutsche Forschungsgemeinschaft (DFG, German Research Foundation) through the CRC-TR 211 `Strong-interaction matter under extreme conditions' -- project number 315477589 -- TRR 211.

\clearpage

\appendix

\section*{Appendix: Fit parameter tables}

Here we provide the data for the fits shown in Figs.\ref{fig:FM_Higgs_deconf_fits}, \ref{fig:FM_deconf_Higgs_fits}~and~\ref{fig:FM_conf_deconf_fit}. 

\subsection*{Higgs phase: $\beta=1.0$, $\gamma=0.228$}

The integrated autocorrelation time for both expectation values entering the definition of FM operators at all distances lie within the range $1.0$ -- $1.2$.
In the analysis we use bin size 50, that is significantly larger than the autocorrelation time. 

We fit the numerator (Table~\ref{tbl:fit-num-b1-y0228}) to 
\begin{equation*}
\left\langle V(R) \right\rangle^2 = A_V \exp \left(-\sigma_V R\right) + B_V \exp \left(-\sigma_{V,2} R\right) \ ,
\end{equation*}
the denominator (Table~\ref{tbl:fit-den-b1-y0228}) to 
\begin{equation*}
\left\langle W(R) \right\rangle = A_W \exp \left(-\sigma_W R\right) \ ,
\end{equation*}
and the ratio (Table~\ref{tbl:fit-b1-y0228}) to 
\begin{equation*}
H(R, T) = A \exp \left(-\sigma R\right) + C \ .
\end{equation*}

\begin{table}[h!]
\centering
 \begin{tabular}{||c c c c c c c c||} 
 \hline
 Operator & $R$ & $A_V$ & $\sigma_V$ & $B_V$ & $\sigma_{V,2}$ & $\chi^2_r$ & p \\ 
 \hline\hline
 spatial & 8 -- 32 & 0.061847(15) & 0.0032931(35) & 0.0318(17) & 0.504(7) & 1.22 & 0.28 \\ 
    & 10 -- 32 & 0.061843(15) & 0.003291(5) & 0.027(5) & 0.488(21) & 1.27 & 0.25 \\ 
    & 12 -- 32 & 0.061857(15) & 0.003298(4) & -- & -- & 0.67 & 0.69 \\ 
    & 10 -- 30 & 0.061841(17) & 0.003287(6) & 0.024(7) & 0.471(31) & 1.25 & 0.27 \\ 
\hline
$T=4$ & 8 -- 32 & 0.061449(15) & 0.001651(4) & 0.0316(18) & 0.503(7) & 1.13 & 0.33 \\
      & 10 -- 32 & 0.061444(15) & 0.001649(4) & 0.025(4) & 0.479(19) & 1.16 & 0.32 \\
      & 12 -- 32 & 0.061558(15) & 0.001656(4) & -- & -- & 
      0.67 & 0.69 \\ 
      & 10 -- 30 & 0.061443(16) & 0.001647(5) & 0.024(5) & 0.476(21) & 
      1.30 & 0.25 \\ 
\hline
$T=6$ & 8 -- 32 & 0.061247(15) & 0.0016517(35) & 0.0314(17) & 0.503(7) & 1.11 & 0.35 \\
      & 10 -- 32 & 0.061244(15) & 0.001650(4) & 0.0258(30) & 0.483(13) & 1.17 & 0.31 \\
      & 12 -- 32 & 0.061257(15) & 0.001657(4) & -- & -- & 0.68 & 0.69 \\
      & 10 -- 30 & 0.061243(15) & 0.001648(5) & 0.025(5) & 0.480(21) & 1.31 & 0.24 \\
\hline
$T=8$ & 8 -- 32 & 0.061044(15) & 0.0016503(34) & 0.0312(17) & 0.503(7) & 1.22 & 0.27 \\
      & 10 -- 32 & 0.061040(15) & 0.001649(4) & 0.0281(4) & 0.492(15) & 1.31 & 0.23 \\
      & 12 -- 32 & 0.061054(15) & 0.001655(4) & -- & -- & 
      0.83 & 0.56 \\
      & 10 -- 30 & 0.061039(16) & 0.001646(5) & 0.0248(5) & 0.478(24) & 1.39 & 0.21 \\
 \hline
 \end{tabular}
 \caption{Fit parameters for the fits of the numerators of the FM operators at $\beta=1.0$, $\gamma=0.228$.}
 \label{tbl:fit-num-b1-y0228}
\end{table}

\begin{table}[h!]
\centering
 \begin{tabular}{||c c c c c c ||} 
 \hline
 Operator & $R$ & $A_W$ & $\sigma_W$ & $\chi^2_r$ & p \\ 
 \hline\hline
 spatial & 8 -- 32 & 1.0002380(8) & 0.0032836(8) & 1.35 & 0.19 \\
         & 10 -- 32 & 1.0002373(12) & 0.0032835(8) & 1.45 & 0.15 \\
         & 12 -- 32 & 1.0002366(15) & 0.0032835(8) & 1.59 & 0.11 \\
         & 10 -- 30 & 1.0002372(12) & 0.0032836(8) & 1.59 & 0.11 \\
\hline
$T=4$   & 8 -- 32 & 0.9936944(20) & 0.0016417(4) & 0.34 & 0.98 \\
        & 10 -- 32 & 0.9936944(20) & 0.0016417(4) & 0.37 & 0.96 \\
        & 12 -- 32 & 0.9936944(21) & 0.0016417(4) & 0.41 & 0.93 \\
        & 10 -- 30 & 0.9936943(20) & 0.0016417(4) & 0.32 & 0.97 \\
\hline
$T=6$  & 8 -- 32 & 0.9904398(31) & 0.0016417(4) & 0.83 & 0.61 \\
       & 10 -- 32 & 0.9904397(31) & 0.0016417(4) & 0.83 & 0.60 \\
       & 12 -- 32 & 0.9904397(31) & 0.0016417(4) & 0.93 & 0.50 \\
       & 10 -- 30 & 0.9904393(31) & 0.0016417(4) & 0.62 & 0.78 \\
\hline
$T=8$  & 8 -- 32 & 0.987196(4) & 0.0016417(4) & 1.09 & 0.37 \\
       & 10 -- 32 & 0.987195(4) & 0.0016417(4) & 1.17 & 0.30 \\
       & 12 -- 32 & 0.987196(4) & 0.0016417(4) & 1.29 & 0.24 \\
       & 10 -- 30 & 0.987195(4) & 0.0016417(4) & 1.12 & 0.35 \\
 \hline
 \end{tabular}
 \caption{Fit parameters for the fits of the denominators of the FM operators at $\beta=1.0$, $\gamma=0.228$.}
 \label{tbl:fit-den-b1-y0228}
\end{table}

\begin{table}[h!]
\centering
 \begin{tabular}{||c c c c c c c||} 
 \hline
 Operator & $R$ & $A$ & $\sigma$ & $C$ & $\chi^2_r$ & p \\ 
 \hline\hline
 spatial   & 8 -- 32 & 0.0290(12) & 0.488(5) & 0.061819(14) & 1.45 & 0.15 \\
           & 10 -- 32 & 0.0213(24) & 0.458(11) & 0.061821(14) & 1.23 & 0.27 \\
           & 12 -- 32 & 0.045(13) & 0.520(25) & 0.061821(14) & 1.16 & 0.32 \\
          & 10 -- 30 & 0.0215(25) & 0.459(12) & 0.061822(14) & 1.12 & 0.34 \\
\hline
$T=4$   & 8 -- 32 & 0.0290(12) & 0.488(5) & 0.061825(14) & 1.38 & 0.18 \\
        & 10 -- 32 & 0.0212(25) & 0.457(12) & 0.061825(14) & 1.15 & 0.32 \\
        & 12 -- 32 & 0.040(8) & 0.510(17) & 0.061826(14) & 1.14 & 0.33 \\
        & 10 -- 30 & 0.0213(24) & 0.458(12) & 0.061826(14) & 1.21 & 0.29 \\
\hline
$T=6$   & 8 -- 32 & 0.0288(12) & 0.488(5) & 0.061824(14) & 1.40 & 0.17 \\
        & 10 -- 32 & 0.0212(30) & 0.458(15) & 0.061823(14) & 1.20 & 0.29 \\
        & 12 -- 32 & 0.038(9) & 0.507(20) & 0.061826(14) & 1.23 & 0.28 \\
        & 10 -- 30 & 0.0213(24) & 0.458(12) & 0.061825(14) & 1.27 & 0.26 \\
\hline
$T=8$   & 8 -- 32 & 0.0291(12) & 0.489(5) & 0.061822(14) & 1.41 & 0.17 \\
        & 10 -- 32 & 0.0222(26) & 0.462(12) & 0.061823(14) & 1.28 & 0.24 \\
        & 12 -- 32 & 0.046(20) & 0.52(4) & 0.061824(14) & 1.26 & 0.26 \\
        & 10 -- 30 & 0.0224(25) & 0.463(11) & 0.061824(14) & 1.26 & 0.26 \\
 \hline
 \end{tabular}
 \caption{Fit parameters for the fits of the FM operators at $\beta=1.0$, $\gamma=0.228$.}
 \label{tbl:fit-b1-y0228}
\end{table}

\subsection*{Deconfinement phase: $\beta=1.0$, $\gamma=0.2$}

The integrated autocorrelation time for both expectation values at all distances lie within the range $0.98$ -- $1.02$.
In the analysis we use bin size 50, that is significantly larger than the autocorrelation time. 

We fit the numerator (Table~\ref{tbl:fit-num-b1-y02}) to 
\begin{equation*}
\left\langle V(R) \right\rangle^2 = A_V \exp \left(-\sigma_V R\right) + B_V \exp \left(-\sigma_{V,2} R\right) \ ,
\end{equation*}
the denominator (Table~\ref{tbl:fit-den-b1-y02}) to 
\begin{equation*}
\left\langle W(R) \right\rangle = A_W \exp \left(-\sigma_W R\right) \ ,
\end{equation*}
and the ratio (Table~\ref{tbl:fit-b1-y02}) to 
\begin{equation*}
H(R, T) = A \exp \left(-\sigma R\right) \ .
\end{equation*}

\begin{table}[h!]
\centering
 \begin{tabular}{||c c c c c c c c||} 
 \hline
 Operator & $R$ & $A_V$ & $\sigma_V$ & $B_V$ & $\sigma_{V,2}$ & $\chi^2_r$ & p \\ 
 \hline\hline
 spatial   & 2 -- 10 & 0.0175(7) & 1.240(6) & 0.365(5) & 2.139(11) & 0.32 & 0.57 \\
           & 2 -- 12 & 0.0174(7) & 1.239(6) & 0.365(5) & 2.139(11) & 0.20 & 0.82 \\
           & 4 -- 12 & 0.05634(26) & 1.4197(12) & -- & -- & 445.10 & 0.00 \\
           & 4 -- 14 & 0.02225(19) & 1.2714(14) & -- & -- & 17.84 & 0.00 \\
\hline
$T=8$ & 2 -- 10 & 0.0170(7) & 1.236(6) & 0.358(5) & 2.134(11) & 0.19 & 0.66 \\
      & 2 -- 12 & 0.0170(7) & 1.236(6) & 0.358(5) & 2.134(11) & 0.10 & 0.91 \\
      & 4 -- 12 & 0.0153(30) & 1.224(22) & 0.25(13) & 2.03(17) & 0.03 & 0.87 \\
      & 4 -- 14 & 0.02107(17) & 1.2616(13) & 1.540(32) & 2.556(6) & 0.25 & 0.78 \\
 \hline
 \end{tabular}
 \caption{Fit parameters for the fits of the numerators of the FM operators at $\beta=1.0$, $\gamma=0.2$.}
 \label{tbl:fit-num-b1-y02}
\end{table}

\begin{table}[h!]
\centering
 \begin{tabular}{||c c c c c c ||} 
 \hline
 Operator & $R$ & $A_W$ & $\sigma_W$ & $\chi^2_r$ & p \\ 
 \hline\hline
 spatial & 4 -- 32 & 1.00028122(30) & 0.0036428(6) & 1.09 & 0.36 \\
         & 6 -- 32 & 1.0002813(5) & 0.0036428(6) & 1.18 & 0.29 \\
         & 8 -- 32 & 1.0002797(7) & 0.0036427(6) & 0.84 & 0.60 \\
         & 4 -- 30 & 1.00028123(30) & 0.0036428(6) & 1.17 & 0.30 \\
\hline
$T=8$   & 4 -- 32 & 0.9858079(31) & 0.00182135(31) & 0.36 & 0.98 \\
        & 6 -- 32 & 0.9858079(31) & 0.00182135(31) & 0.39 & 0.97 \\
        & 8 -- 32 & 0.9858080(31) & 0.00182136(31) & 0.43 & 0.95 \\
        & 4 -- 30 & 0.9858079(31) & 0.00182135(31) & 0.33 & 0.98 \\
 \hline
 \end{tabular}
 \caption{Fit parameters for the fits of the denominators of the FM operators at $\beta=1.0$, $\gamma=0.2$.}
 \label{tbl:fit-den-b1-y02}
\end{table}

\begin{table}[h!]
\centering
 \begin{tabular}{||c c c c c c||} 
 \hline
 Operator & $R$ & $A$ & $\sigma$ & $\chi^2_r$ & p \\ 
 \hline\hline
 spatial   & 6 -- 12 & 0.0238(6) & 1.272(4) & 2.11 & 0.12 \\
  & 6 -- 14 & 0.02406(5) & 1.2744(4) & 1.33 & 0.26 \\
  & 6 -- 16 & 0.02399(5) & 1.2740(4) & 11.09 & 0.00 \\
  & 8 -- 14 & 0.0171(6) & 1.232(4) & 0.00 & 1.00 \\
\hline
$T=8$   & 6 -- 12 & 0.0235(6) & 1.271(4) & 1.80 & 0.17 \\
  & 6 -- 14 & 0.02382(5) & 1.2728(4) & 1.20 & 0.31 \\
  & 6 -- 16 & 0.02376(5) & 1.2723(4) & 10.45 & 0.00 \\
  & 8 -- 14 & 0.0172(6) & 1.232(4) & 0.00 & 1.00 \\
 \hline
 \end{tabular}
 \caption{Fit parameters for the fits of the FM operators at $\beta=1.0$, $\gamma=0.2$.}
 \label{tbl:fit-b1-y02}
\end{table}

\subsection*{Confinement phase: $\beta=0.75$, $\gamma=0.21$}

The integrated autocorrelation time for both expectation values at all distances lie within the range $0.99$ -- $13.0$.
In the analysis we use bin size 120, that is significantly larger than the autocorrelation time. 

We fit the numerator (Table~\ref{tbl:fit-num-b075-y021}) to 
\begin{equation*}
\left\langle V(R) \right\rangle^2 = A_V \exp \left(-\sigma_V R\right) + B_V \exp \left(-\sigma_{V,2} R\right) \ ,
\end{equation*}
the denominator (Table~\ref{tbl:fit-den-b075-y021}) to 
\begin{equation*}
\left\langle W(R) \right\rangle = A_W \exp \left(-\sigma_W R\right) + B_W \exp \left(-\sigma_{W,2} R T - \alpha R \right) \ ,
\end{equation*}
for the spatial Wilson loop and 
\begin{equation*}
\left\langle W(R) \right\rangle = A_W \exp \left(-\sigma_W R\right) + B_W \exp \left(-\sigma_{W,2} R\right) \ ,
\end{equation*}
for the temporal one (where we cannot distinguish between $\alpha$ and $\sigma_{W,2}$ contributions), 
and the ratio (Table~\ref{tbl:fit-b075-y021}) to 
\begin{equation*}
H(R, T) = A \exp \left(-\sigma R\right) \ .
\end{equation*}

\begin{table}[h!]
\centering
 \begin{tabular}{||c c c c c c c c||} 
 \hline
 Operator & $R$ & $A_V$ & $\sigma_V$ & $B_V$ & $\sigma_{V,2}$ & $\chi^2_r$ & p \\ 
 \hline\hline
 spatial & 2 -- 18 & 0.0345(4) & 1.3883(16) & 0.4237(23) & 2.253(5) & 7.36 & 0.00 \\
  & 4 -- 18 & 0.0287(8) & 1.3704(29) & 0.195(12) & 1.998(23) & 0.20 & 0.94 \\
  & 6 -- 18 & 0.0283(23) & 1.369(7) & 0.17(11) & 1.97(15) & 0.26 & 0.85 \\
  & 4 -- 16 & 0.0286(8) & 1.3702(30) & 0.195(12) & 1.997(23) & 0.06 & 0.98 \\
  \hline
 $T=4$ & 6 -- 22 & 0.00450(35) & 1.098(6) & 0.0426(21) & 1.489(16) & 1.90 & 0.09 \\
  & 8 -- 22 & 0.0027(6) & 1.068(14) & 0.0249(29) & 1.364(33) & 1.37 & 0.24 \\
  & 10 -- 22 & 0.0022(11) & 1.059(24) & 0.020(7) & 1.33(7) & 1.79 & 0.15 \\
  & 8 -- 20 & 0.0023(6) & 1.059(16) & 0.0235(25) & 1.346(31) & 0.51 & 0.68 \\
  \hline
$T=6$ & 6 -- 22 & 0.00346(29) & 1.105(6) & 0.0372(20) & 1.508(17) & 1.58 & 0.16 \\
  & 8 -- 22 & 0.0025(5) & 1.087(13) & 0.024(4) & 1.42(4) & 1.52 & 0.19 \\
  & 10 -- 22 & 0.0032(7) & 1.098(14) & 0.07(10) & 1.56(18) & 1.81 & 0.14 \\
  & 8 -- 20 & 0.0023(5) & 1.082(14) & 0.023(4) & 1.40(4) & 1.47 & 0.22 \\
\hline
$ T=8$  & 6 -- 22 & 0.00269(26) & 1.115(7) & 0.0316(19) & 1.521(19) & 1.23 & 0.29 \\
  & 8 -- 22 & 0.0017(5) & 1.090(17) & 0.019(4) & 1.41(5) & 1.10 & 0.36 \\
  & 10 -- 22 & 0.0014(8) & 1.078(31) & 0.014(7) & 1.35(10) & 1.41 & 0.24 \\
  & 8 -- 20 & 0.0017(5) & 1.088(18) & 0.0186(34) & 1.41(5) & 1.38 & 0.25 \\
 \hline
 \end{tabular}
 \caption{Fit parameters for the fits of the numerators of the FM operators at $\beta=0.75$, $\gamma=0.21$.}
 \label{tbl:fit-num-b075-y021}
\end{table}

\begin{table}[h!]
\centering
 \begin{tabular}{||c c c c c c c c ||} 
 \hline
 Operator & $R$ & $A_W$ & $\sigma_W$ & $B_W$ & $\sigma_{W,2}$ & 
 $\chi^2_r$ & p \\ 
 \hline\hline
 spatial   & 12 -- 32 & 0.19(8) & 0.347(13) & 1.20(5) & 0.00705(12) & 
 2.15 & 0.04 \\
  & 14 -- 32 & 0.24(17) & 0.358(21) & 1.22(10) & 0.00691(20) & 
  1.90 & 0.09 \\
  & 14 -- 30 & 0.24(12) & 0.358(15) & 1.22(7) & 0.00690(16) & 
  2.38 & 0.05 \\
  \hline 
  $T=4$  & 8 -- 32 & 0.77336(15) & 0.109447(33) & 0.03968(16) & 0.3482(19) & 54.43 & 0.00 \\
  & 10 -- 32 & 0.77049(26) & 0.10921(4) & 0.03321(22) & 0.2942(28) & 35.63 & 0.00 \\
  & 12 -- 32 & 0.7663(6) & 0.10897(5) & 0.02870(32) & 0.239(4) & 18.75 & 0.00 \\
  & 10 -- 30 & 0.77035(22) & 0.109185(34) & 0.03336(22) & 0.2937(24) & 31.93 & 0.00 \\
  \hline
  $T=6$  & 8 -- 32 & 0.64693(31) & 0.12780(6) & 0.06459(26) & 0.3521(20) & 95.98 & 0.00 \\
  & 10 -- 32 & 0.6361(7) & 0.12683(7) & 0.0566(4) & 0.2742(27) & 63.27 & 0.00 \\
  & 12 -- 32 & 0.6172(19) & 0.12578(11) & 0.0624(15) & 0.2144(29) & 35.45 & 0.00 \\
  & 10 -- 30 & 0.6357(7) & 0.12675(7) & 0.0571(4) & 0.2741(25) & 52.30 & 0.00 \\
  \hline
  $T=8$  & 8 -- 32 & 0.5299(5) & 0.14234(10) & 0.08901(33) & 0.3512(22) & 158.32 & 0.00 \\
  & 10 -- 32 & 0.4912(16) & 0.13864(15) & 0.0979(13) & 0.2450(18) & 106.75 & 0.00 \\
  & 12 -- 32 & 0.415(5) & 0.13449(28) & 0.158(5) & 0.1916(16) & 55.73 & 0.00 \\
  & 10 -- 30 & 0.4922(17) & 0.13857(16) & 0.0976(14) & 0.2473(20) & 89.92 & 0.00 \\
 \hline
 \end{tabular}
 \caption{Fit parameters for the fits of the denominators of the FM operators at $\beta=0.75$, $\gamma=0.21$.}
 \label{tbl:fit-den-b075-y021}
\end{table}

\begin{table}[h!]
\centering
 \begin{tabular}{||c c c c c c||} 
 \hline
 Operator & $R$ & $A$ & $\sigma$ & $\chi^2_r$ & p \\ 
 \hline\hline
 spatial   & 10 -- 20 & 0.010141(17) & 1.02625(17) & 7.10 & 0.00 \\
  & 12 -- 20 & 0.00736(11) & 0.9995(12) & 0.36 & 0.78 \\
  & 14 -- 20 & 0.0045(5) & 0.964(8) & 0.02 & 0.98 \\
  & 12 -- 18 & 0.0071(8) & 0.996(10) & 0.51 & 0.60 \\
  \hline
 $T=4$   & 10 -- 20 & 0.01086(22) & 1.0340(21) & 10.30 & 0.00 \\
  & 12 -- 20 & 0.00780(35) & 1.007(4) & 3.89 & 0.01 \\
  & 14 -- 20 & 0.0051(5) & 0.977(7) & 0.54 & 0.58 \\
  & 12 -- 18 & 0.00780(35) & 1.007(4) & 5.80 & 0.00 \\
  \hline
 $T=6$   & 10 -- 20 & 0.00989(22) & 1.0230(24) & 8.73 & 0.00 \\
  & 12 -- 20 & 0.00682(35) & 0.992(4) & 2.43 & 0.06 \\
  & 14 -- 20 & 0.0050(5) & 0.971(8) & 1.38 & 0.25 \\
  & 12 -- 18 & 0.00677(35) & 0.992(4) & 2.79 & 0.06 \\
  \hline
 $T=8$   & 10 -- 20 & 0.00947(25) & 1.0200(28) & 6.10 & 0.00 \\
  & 12 -- 20 & 0.0068(4) & 0.993(5) & 3.16 & 0.02 \\
  & 14 -- 20 & 0.0041(5) & 0.957(9) & 0.42 & 0.65 \\
  & 12 -- 18 & 0.0068(4) & 0.993(5) & 4.73 & 0.01 \\
 \hline
 \end{tabular}
 \caption{Fit parameters for the fits of the FM operators at $\beta=0.75$, $\gamma=0.21$.}
 \label{tbl:fit-b075-y021}
\end{table}

\end{document}